\documentclass[article]{aa} 
\usepackage{multirow} 
\usepackage{natbib}
\usepackage{amssymb,amsmath}
\usepackage{graphicx, subfigure}
\usepackage{siunitx}
\usepackage{tablefootnote}
\usepackage{url}
\usepackage{threeparttable}
\usepackage[table]{xcolor}
\usepackage{makecell}
\usepackage{epstopdf}
\usepackage{txfonts}
\usepackage{ textcomp }

\usepackage{color, colortbl}
\usepackage{booktabs}
\usepackage{placeins}
\usepackage[export]{adjustbox}
\usepackage{hyperref} 
\definecolor{Gray}{gray}{0.9}

\begin{document}

\newcommand{\ra}[4]{#1^{\rm h}#2^{\rm m}#3^{\rm s}.#4}
\newcommand{\dec}[4]{#1^{\rm \circ}#2\arcmin#3\farcs#4}
\newcommand{\jtwoone}{\mbox{$J$=2$-$1}} 
\newcommand{\jonezero}{\mbox{$J$=1$-$0}} 
   \title{ALMA observations of CO isotopologues towards six obscured post-AGB stars}

   \author{T. Khouri\inst{1}\thanks{{\it Send offprint requests to T. Khouri}
   \newline \email{theokhouri@gmail.com}}, D. Tafoya\inst{1}, W. H. T. Vlemmings\inst{1}, H. Olofsson\inst{1},
   C. S\'anchez Contreras\inst{2}, J. Alcolea\inst{3}, J. F. G\'omez\inst{4}, L. Velilla-Prieto\inst{5}, R. Sahai\inst{6},
   M. Santander-Garc\'ia\inst{3}, V. Bujarrabal\inst{7}, A. Karakas\inst{8,9},
   M. Saberi\inst{10,11}, I. Gallardo Cava\inst{3}, H. Imai\inst{12,13}, A. F. P\'erez-S\'anchez\inst{14}
}

\institute{Department of Space, Earth and Environment, Chalmers University of Technology, Onsala Space Observatory, 439 92 Onsala, Sweden 
\and Centro de Astrobiologia (CAB), CSIC-INTA, ESAC Campus, Camino Bajo del Castillo s/n, 28692 Villanueva de la Ca\~nada, Madrid, Spain 
\and Observatorio Astron\'omico Nacional (OAN-IGN), Alfonso XII No. 3, 28014 Madrid, Spain 
\and Instituto de Astrof\'{\i}sica de Andaluc\'{\i}a, CSIC, Glorieta de la Astronom\'{\i}a s/n, E-18008 Granada, Spain 
\and Instituto de Fisica Fundamental (CSIC), C/Serrano,123, 28006 Madrid, Spain 
\and Jet Propulsion Laboratory, MS 183-900, California Institute of Technology, Pasadena, CA 91109, USA 
\and Observatorio Astron\'omico Nacional (OAN-IGN), Ap112, 28803 Alcal\'a de Henares, Madrid, Spain 
\and School of Physics and Astronomy, Monash University, VIC 3800, Australia 
\and ARC Centre of Excellence for All Sky Astrophysics in 3 Dimensions (ASTRO 3D), Australia 
\and Rosseland Centre for Solar Physics, University of Oslo, PO Box 1029, Blindern 0315, Oslo, Norway 
\and Institute of Theoretical Astrophysics, University of Oslo, P.O. Box 1029 Blindern, 0315 Oslo, Norway 
\and Center for General Education, Institute for Comprehensive Education, Kagoshima University, 1-21-30 Korimoto, Kagoshima 890-0065, Japan 
\and Amanogawa Galaxy Astronomy Research Center (AGARC), Graduate School of Science and Engineering,
Kagoshima University, 1-21-30 Korimoto, Kagoshima 890-0065, Japan 
\and Joint ALMA Observatory, Avenida Alonso de C\'ordova 3107, Vitacura, Santiago, Chile 
}

  \abstract
  {Low- and intermediate-mass stars evolve through the asymptotic giant branch (AGB), when an efficient mass-loss process removes
  a significant fraction of their initial mass. For most sources, this mass-loss process relies on the interplay between convection, stellar pulsations,
  and dust formation. However, predicting the mass-loss history of a given star from first principles is complex and still not possible. At the end of
  the AGB at least some stars experience a substantial increase in their mass-loss rate for unknown reasons, creating
  post-AGB objects that are completely enshrouded in thick dusty envelopes. Recent studies have suggested that some of these sources may
  be the product of interactions between an evolved star with a close companion.}
  {We observed six obscured post-AGB stars (four C-rich and two O-rich sources)
  to constrain the properties of their circumstellar envelopes, recent mass-loss histories,
  and initial mass of the central stars.}
  {We use observations of the \jtwoone\, line of $^{13}$CO, C$^{17}$O, and C$^{18}$O with the Atacama Large Millimeter / submillimeter Array (ALMA)
  to determine the circumstellar gas masses and the $^{17}$O/$^{18}$O isotopic ratios, which
  correlate with initial mass. These are interpreted based on results from stellar evolution models in the
  literature and existing observations of other post-AGB stars.}
  {
  Based on the inferred  $^{17}$O/$^{18}$O isotopic ratios, we find all stars to have relatively low initial masses ($< 2~M_\odot$)
 contrary to suggestions in the literature of higher masses for some sources. One of the C-rich sources, HD~187885, has a low $^{17}$O/$^{18}$O ratio,
 which together with a low metallicity implies a relatively low mass ($\sim 1.15~M_\odot$) for a carbon star. For all but one source (GLMP~950),
 we observe kinematic components with velocities $\gtrsim 30$~km~s$^{-1}$, which are faster than typical AGB wind expansion velocities.
 For most sources, these higher-velocity outflows display
 point-symmetric morphologies. The case of Hen~3-1475 is particularly spectacular, with the high-velocity molecular outflow appearing
 interleaved with the high-velocity outflow of ionised gas observed at optical wavelengths. Based on the size of the emission regions of the
 slow components of the outflows, we derive typical kinematic ages associated with the C$^{18}$O~$J=2-1$ emission $\lesssim 1500$~years and
 obtain relatively high associated mass-loss rates ($\gtrsim10^{-4}~M_\odot~{\rm yr}^{-1}$). The sources with known spectral types are found
 to have evolved faster than expected based on stellar evolutionary models.}
{}%
               
\titlerunning{}
\authorrunning{T. Khouri et al.}

\maketitle
%

\section{Introduction}

Stars with initial masses between roughly 1 and 8~$M_\odot$ evolve through the the asymptotic giant branch (AGB) at the end of
their lives \citep{Karakas2014}.
In this phase, nuclear burning of helium and hydrogen happens in separate layers around the inert core rich in
carbon and oxygen \citep{Herwig2005}. The dense, stratified inner region is surrounded by a
large ($\sim 1$~au) convective stellar envelope, making AGB stars large red giants.
The combination of large sizes, low values of surface gravity, and cool and pulsating outer layers leads to such a strong mass-loss process on the AGB phase
\citep[between $10^{-7}$ and $10^{-4}~M_\odot~{\rm yr}^{-1}$,][]{Hoefner2018} that it dominates the evolution of the star from its onset.
This process continues until the stellar envelope has essentially been stripped away. The high-energy radiation
field from the newly exposed core eventually ionises the ejected gas. If the timing between outflow expansion and
evolution of the central remnant is in the right range, the envelope will be visible as a planetary nebula \citep[PN, ][]{Kwok1994}.

Explaining the shapes and properties of planetary nebulae (PNe) has been a long-standing problem in astrophysics.
The transition from the mostly spherically symmetric outflows of AGB stars to the complex morphologies of
PNe \citep[e.g., ][]{Sahai1998,Parker2006,Sahai2007,Sahai2011} has been a particularly challenging problem to solve. Recent observations of AGB
stars have revealed distributions of the circumstellar molecular gas \citep{Decin2020}
which are reminiscent of complex patterns of arcs and shells seen in the haloes of some
PNe. These structures have been suggested to be created by stellar or even planetary companions.
The ejection of clumps of gas in random directions due to convection has also been proposed as an important shaping mechanism
of AGB outflows \citep{Velilla-Prieto2023}.
However, the evolution of the structures of the density distributions into later evolutionary phases, their prevalence in the envelopes of AGB stars,
and the average fraction of the circumstellar envelope (CSE) mass contained in such structures remains not well constrained. On large scales
and at lower resolutions, the molecular CSEs of AGB stars appear spherically symmetric on average \citep{Ramstedt2020}.

One important process invoked to explain the circumstellar masses of PNe is the
so-called super-wind, which has been proposed to be a short phase of intense mass loss (with rates $\sim 10^{-4}~M_\odot~{\rm yr}^{-1}$)
taking place at the end of the AGB \citep{Renzini1981,Vassiliadis1993}. One possibility
is that PNe are mostly the outcome of binary interactions, in which a type of common-envelope phase takes place
\citep[for an overview, see ][and references therein]{Jones2017,Boffin2019}.
However, only a relatively small fraction ($\sim 20\%$) of stars evolving through the AGB
are expected to have stellar companions at small-enough separations to {potentially} induce this type of evolution \citep[e.g., ][]{Boffin2019,Khouri2022}.
{ This relatively low fraction might not be in tension with the observed number of PNe
because population synthesis calculations indicate that only a fraction of low- and intermediate-mass
stars produce bright PNe \citep{Moe2006,DeMarco2009}. Moreover, single-star models do not maintain the required
rotation rates to reproduce the observed fraction of non-symmetric PNe \citep[e.g.][]{GarciaSegura2014}. Finally, the empirical binarity fraction of
central-stars of PNe is only a lower limit $\gtrsim 20~\%$ \citep{Jones2017} to the actual fraction, given the observational challenges for this type of study \citep[e.g.][]{DeMarco2015}.
Nonetheles, the clear correlation between the orientation of the PNe axis and the axis of the orbital plane in PNe with known binary central stars \citep{Hillwig2016} implies that companions
strongly affect the shape of PNe.}

{ Given the observational evidence and the current limitations to studies of PNe and AGB stars},
studying the phase before the PN stage and immediately after the AGB, { the post-AGB phase, is crucial for understanding the final stages of the lives of low- and intermediate-mass stars.}
{ A subclass of post-AGB sources with dusty discs (containing $> 80$ objects)
are known to very often have detectable close companions \citep{vanWinckel2007,Oomen2018}.}
Some of these sources even display gaseous discs with Keplerian rotation that remain stable
over long timescales \citep{vanWinckel2018,GallardoCava2021}.
{ Although determining binary fractions in post-AGB stars more obscured by circumstellar dust is difficult, observations of
the morphology and dynamical components of their CSEs provides important clues. Studies of such objects} have
found high-velocity outflows with excess momentum that cannot be explained by radiation-driven outflows
\citep[e.g., ][]{Bujarrabal2001} and complex morphologies \citep{Sahai2007,Lagadec2011}.
In fact, jets or bipolar outflows are expected to be one of the main agents affecting the morphologies of these systems and of PNe \citep[e.g., ][]{Sahai1998,Sahai2011,Soker2020}.
Circumstellar masses derived based on CO observations of post-AGB sources range broadly from $10^{-2}$ to $1~M_\odot$ \citep{Bujarrabal2001,SanchezContreras2012}. 
The different configurations of binary systems leading to the observed post-AGB morphology and CSE components remain under investigation.

Another important diagnostic is the study of the isotopic ratios of the ejected material, which can be interpreted using
our understanding of nucleosynthesis and stellar evolution. The isotopes of oxygen, and specifically the $^{17}$O/$^{18}$O ratio,
provide particularly powerful constraints on the initial masses of stars in this mass range. For most low- and intermediate-mass stars,
the significant changes to the oxygen isotopic ratios
take place between the main sequence and the red giant branch phases, when convective streams in the stellar envelope reach material processed by nucleosynthesis in the stellar interior in a process referred to as the first dredge up \citep[e.g., ][]{Karakas2014}.
There is overall agreement between different stellar evolution
models on the quantitative values of oxygen isotopic ratios after the first dredge up \citep{DeNutte2017}. However,
the expected dependence of this ratio on the initial stellar mass of stars has not been fully validated by observations thus far \citep{Abia2017}.
The oxygen isotopic ratios have been extensively studied for AGB stars
\citep[e.g., ][]{Harris1987,Smith1990,Hinkle2016,Abia2017,Danilovich2017,DeNutte2017,Lebzelter2019}, but less often for post-AGB objects.

Stellar evolution models also predict whether or not stars with a given initial mass and metallicity
become carbon-rich at the end of the AGB. The increase of the carbon abundance
in excess of that of oxygen in the stellar envelope happens for stars with initial mass between $\sim 1.5$ and $\sim 4$~$M_\odot$
during the third dredge up \citep{Karakas2014}, which is a mixing event that takes place during the AGB phase itself \citep{Iben1983}.
The lower mass limit for stars to become carbon rich is defined by whether the enrichment is efficient and proceeds for long enough, while
the upper mass limit depends on whether the temperature at the base of the convective envelope is high enough to convert carbon mainly into nitrogen
in a process known as Hot Bottom Burning \citep[HBB, e.g.,][]{Herwig2005}. The efficiency of the third dredge up in enriching the stellar envelope
with carbon depends on many parameters, including mixing efficiency, AGB lifetimes, and
prescriptions for the mass-loss rate included in stellar evolution models \citep[e.g., ][]{Stancliffe2004,Abia2017,Rees2024}. 
If HBB takes place, $^{18}$O is expected to be very efficiently destroyed leading to very low $^{18}$O abundances, but the minimum
initial stellar mass for which HBB becomes important is fairly uncertain \citep{Ventura2018}.
A few AGB stars with extreme mass-loss rates seem to belong to this class \citep{Justtanont2012},
while no post-AGB stars with very low $^{18}$O abundances
have been identified thus far to our knowledge.
In fact, recent studies of oxygen isotopic ratios in post-AGB stars \citep{Khouri2022,Alcolea2022} have indicated that some of
the most obscured post-AGB sources
have relatively low initial masses ($\lesssim 3~M_\odot$), contrary to the expectation that these were more massive stars with initial masses $\gtrsim 5~M_\odot$.

A small subclass of post-AGB stars termed water fountains accounts for a significant fraction of the post-AGB objects
with determined oxygen isotopic ratios \citep{Khouri2022}.
These sources show water maser emission with expansion velocities higher than
those of their own OH maser emission, which is not observed in any other class of AGB or post-AGB stars. The water maser emission
is thought to arise from the interaction of a nascent jet with a dense CSE \citep[e.g., ][]{Imai2007}.
The envelopes present tori-like structure oriented mostly perpendicularly to bipolar lobes sculpted by high-velocity jets \citep{Sahai2017,Gomez2018,Tafoya2020}.
The ejection of large amounts of mass in these systems seems to have occurred $\lesssim 100$ years ago \citep[e.g., ][]{Boboltz2007,Day2010}.
The compactness of the emission observed towards water fountains and other highly obscured sources suggests a period of very intense mass loss,
which has been proposed to be triggered by interactions with a nearby companion \citep{Khouri2022,Alcolea2022}. The process
experienced by these sources could be connected to the super-wind phase, but whether these are the same, or related at all, remains presently unknown.

Regarding carbon enrichment, a lot of effort has been put into reproducing the mass range for which stars become carbon rich in stellar evolution models.
Observations of carbon-rich stars in open clusters indicate that the lower limit in initial mass for forming carbon-rich stars is
$\sim~1.5~M_\odot$ \citep{Groenewegen1995} at solar metallicity. A similar limit is obtained by
fitting the luminosity function of carbon stars in the Milky Way
and satellite galaxies using stellar evolution codes \citep[e.g., ][]{Groenewegen2002}, but the initial mass might extend
down to $\sim 1~M_\odot$ at lower metallicities \citep{Stancliffe2005}. However, the predicted minimum luminosity is high compared to observations.
Stellar evolution models are able to reproduce carbon stars with initial masses as low as $\sim 1.5~M_\odot$
by including extra mixing for the third dredge-up to be efficient enough 
\citep[e.g., ][]{Weiss2009,Cristallo2015,Karakas2016,Ventura2018,Goriely2018,Rees2024}.

In this paper, we study a small sample of six obscured
post-AGB objects and investigate the $^{17}$O/$^{18}$O ratio, their recent mass-loss rates, and
the CSE morphology at scales of a few thousand au using the \jtwoone\, transitions of
isotopologues of CO. In Section~\ref{sec:sample} we list the sample stars and their relevant characteristics from literature. In Sections~\ref{sec:obs}
and~\ref{sec:obsRes} we describe the observations and present the results, respectively. The results are discussed in the context of our
current understanding in Section~\ref{sec:disc}. Finally, a summary and conclusions are presented in Section~\ref{sec:conc}.

\section{Sample}
\label{sec:sample}

\subsection{Sample selection}

\begin{figure}[t]
\begin{center}
 \includegraphics[width=0.4\textwidth]{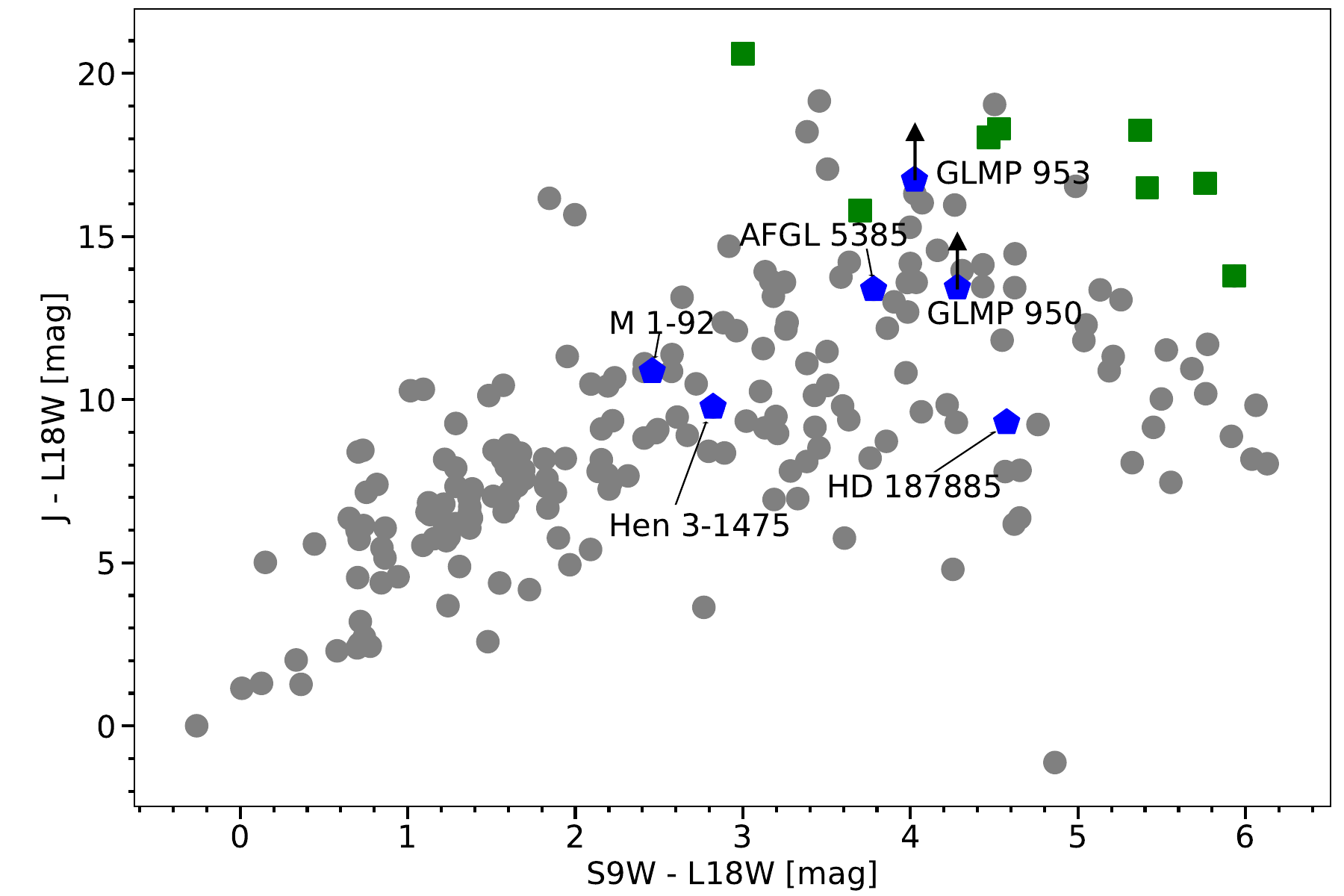}
 \caption{Position of the sources included in our sample in a colour-colour diagram for post-AGB stars. The blue pentagons show the sources
 included in this study, the green squares represent water fountain sources observed by \cite{Khouri2022}, and the grey circles correspond to other known
 post-AGB stars
 catalogued by \cite{Szczerba2007}. The vertical arrows indicate our assessment that the $J-$band magnitude of GLMP~950 and GLMP~953
 is dominated by contaminating sources close to their position on the sky (see Sections~\ref{sec:GLMP950} and \ref{sec:GLMP953} and Figs.~\ref{fig:Spitzer_GLMP950} and \ref{fig:Spitzer_GLMP953}).}
    \label{fig:colour-colour}
\end{center}
\end{figure}

We observed six post-AGB stars selected for having very red colours. To select our sample, we obtained
magnitudes in the $J$ band from 2MASS and in the $S9W$ and $L18W$ bands from AKARI using the catalogue by \cite{Ita2010} for
post-AGB stars listed in \cite{Szczerba2007}.
We selected sources with $J-L18W > 8$ and $S9W-L18W > 2$.
When the magnitudes
for AKARI filters were missing, we obtained values by interpolating the spectral energy distribution (SED) of the given
source using the VizieR photometry tool.  
The colour-colour diagram containing the sample of post-AGB stars is shown in Fig.~\ref{fig:colour-colour}. The sources presented
in this paper are shown by the blue pentagons, while water fountain sources studied in \cite{Khouri2022}  \citep[most of which are not included in the catalogue of ][]{Szczerba2007}
are shown by the green squares. 
Our original sample of post-AGB stars consisted of twenty sources, but only six were successfully observed with ALMA within this project. Four of these
(GLMP~950, GLMP~953, AFGL~5385, and HD~187885) are rich in carbon,
while the other two (Hen~3-1475 and M1-92) have
oxygen-rich composition. { The bluer colors of the two oxygen-rich stars in our sample could in principle be linked to a systematic lower absorption opacity of oxygen-rich
relatively to carbon-rich dust or to lower gas-to-dust ratios in C-rich sources. This is especially intriguing when considering
the similar gas mass-loss rates and timescales derived below. However, the small number of sources in
our sample prevents us from investigating this difference further.}

\subsection{Relevant information on individual sources}

GLMP~950 (IRAS~19454+2920) and GLMP~953 (IRAS~19480+2504) were imaged by
\cite{Lagadec2011} at wavelengths between 8 and 13~$\mu$m using VISIR/Very Large Telescope (VLT).
The authors found GLMP~950 to be spatially resolved in the images,
with a somewhat elliptical shape extending in the east-west direction to a maximum size of $\sim 2 \arcsec$.
GLMP~953 was not spatially resolved in the observations with a full-width at half maximum (FWHM) of the
Gaussian beam of $\sim 0\farcs54$.

AFGL~5385 (IRAS~17441-2411, the Silkworm Nebula) displays  an edge-on torus oriented in the southeast-northwest direction
and bipolar lobes oriented roughly in the north-south direction. An S-shaped structure is also visible within the lobes.
This nebula has been imaged in visible light \citep{Su1998} and in the near-IR \citep{Su2003} with the {\it Hubble Space Telescope (HST)}, and in
the mid-IR with T-ReCS/Gemini South \citep{Volk2007} and VISIR/VLT \citep{Lagadec2011}.
The bipolar lobes have a total extent of $\sim 4\arcsec$, while the torus has a size of $\sim 1\arcsec$.

HD~187885 (IRAS~19500-1709) is commonly classified as a carbon-rich post-AGB star, but its determined elemental abundances indicate
a carbon-to-oxygen ratio close to unity \citep{vanWinckel1996,vanWinckel2000}. The source was also determined to be more metal poor than the
Sun, with [Fe/H] = $-0.6$ \citep{vanWinckel2000}. \cite{Lagadec2011} observed a slightly elliptical source with size $\sim 3\arcsec$ at 11.85 and 12.81~$\mu$m
using VISIR/VLT. They find an inner radius of the CSE of $\sim 0\farcs4$, which confirmed previous indications from
images at 10 and 20~$\mu$m obtained using the OSCIR/Gemini North \citep{Clube2004}. Modelling of the spectral energy distribution (SED)
indicates a dust mass in the CSE of
$\sim 7 \times 10^{-4}~M_\odot$ and an associated duration of the high-mass-loss phase of 4000~yr \citep{Clube2004}, where both
values have been converted to the distance of 2.4~kpc adopted in this work.
As discussed by the authors, under the assumption of a gas-to-dust ratio of 220 these results imply a mass-loss rate of $\sim 4 \times 10^{-4}~M_\odot~{\rm yr}^{-1}$,
which is large in comparison to other determinations also through SED modelling \citep[$4\times10^{-5}~M_\odot~{\rm yr}^{-1}$, ][]{Guertler1996}.
\cite{Bujarrabal2001} found a clear separation between a slow (expansion velocity $\sim 10$~km~s$^{-1}$)
and a faster (expansion velocity $\sim 50$~km~s$^{-1}$)
components of the outflow in CO emission lines.
The authors derived a total circumstellar gas mass of $\sim 7 \times 10^{-3}$ and
$\sim 2.6 \times 10^{-2}~M_\odot$ for the fast and slow outflows, respectively, considering a distance of 1~kpc.
At the distance of 2.4~kpc adopted by us, the fast and slow outflows would contain masses of $\sim 4 \times 10^{-2}$ and
$\sim 0.15~M_\odot$, respectively.

Hen~3-1475 (IRAS~17423-1755) displays a bipolar morphology in images in visible light with a northwest-southeast orientation of the lobes
and atomic line widths which imply a velocity range of $\sim1000$~km~s$^{-1}$ \citep{Bobrowsky1995}. Within the bipolar lobes, a point-symmetric
structure with knots is visible in emission of atomic lines \citep{Borkowski1997}.
Hen~3-1475 is the only post-AGB object where an ultrafast ($\sim 2300$~km~s$^{-1}$)
collimated jet, presumably in a pristine stage before interacting with the surrounding nebula,
has been identified with HST/STIS \citep{SanchezContreras2001}.
 The inclination of the ultrafast wind is $\sim75^\circ$ with respect to the plane of the sky, whereas the mean inclination of the
 nebula (reflecting lobes and CO hourglass) is $\sim 50^\circ$.
\cite{Borkowski2001} determined a distance of 8.3~kpc to Hen~3-1475 based on proper
motions of the jet knots, implying a luminosity of $25000~L_\odot$. This led them to conclude that Hen~3-1475 is a massive
AGB star or the product of binary evolution. \cite{Riera2003} finds a smaller distance ($5.8 \pm 0.9$~kpc), but also suggests a high
initial mass based on a nitrogen overabundance reported by \cite{Riera1995}. A distance of 5.8~kpc implies a luminosity of $\sim 12000~{\rm L}_\odot$.
\cite{Bujarrabal2001} identified a slow (with expansion velocity $\sim 7$~km~s$^{-1}$)
and a faster (with expansion velocity $\sim 50$~km~s$^{-1}$) component in
the outflow of Hen~3-1475 based on CO and $^{13}$CO~\jonezero\,
and \jtwoone\, lines. They inferred
a mass of 0.16~$M_\odot$ and 0.47~$M_\odot$ for the two dynamical components, respectively, for a distance of
5~kpc.
\cite{Huggins2004} derived a circumstellar mass of 0.64~$M_\odot$ based on the
measured continuum emission of $5.3 \pm 1.2$  and $31 \pm 4$~mJy at 2.6 and 1.3~mm, respectively, for a distance of 5.8~kpc
and a gas-to-dust ratio of 100. The CO~\jonezero\, and \jtwoone\,  lines
were most likely affected by optical depth effects and implied a lower limit for the circumstellar mass of 0.19~$M_\odot$.
\cite{Manteiga2011} suggest Hen~3-1475 to be a high-mass ($M > 3~M_\odot$) O-rich post-AGB star based on the presence of water ice
absorption.
We adopt a distance of 5.8~kpc to Hen~3-1475, following \cite{Riera1995}.

M1-92 (IRAS 19343+2926, Minkowski's Footprint) displays a bipolar shape at
optical wavelengths \citep{Trammell1996}. The strong molecular emission has been extensively observed and modelled in detail 
\citep{Bujarrabal1997,Alcolea2007,Alcolea2022}. The lines of the isotopologues of CO have been studied previously. In particular,
emission from C$^{17}$O and C$^{18}$O have been used to determine a $^{17}$O/$^{18}$O ratio of $1.6\pm0.15$ \citep{Alcolea2022}. Given these detailed studies,
we refrain from discussing the morphology of this source extensively, and only report the values measured by us highlighting differences to
values in the literature.

We adopt distances to each object from \cite{Vickers2015}, with the exception of Hen~3-1475, for which we adopt the value from \cite{Riera1995}.
The distances used in this work, and their respective references, are provided in Table~\ref{tab:cont}.

\begin{table}
\centering
\caption{Parameters of the observations for each source.}
\label{tab:obs}
\begin{tabular}{ l @{\phantom{aa}}c @{\phantom{aa}}c c }
Source & Beam & Noise level & MRS \\
 & [FWHM and PA] & [mJy/beam] & [$\arcsec$]\\
\hline
 GLMP~950 & $1\farcs09 \times 0\farcs77$ / \phantom{0}-6$^\circ$ & 2.5 & 10.4 \\
 GLMP~953 & $1\farcs00 \times 0\farcs78$ / \phantom{0}-8$^\circ$ & 2.5 & 9.8 \\
 AFGL~5385 & $1\farcs19 \times 0\farcs63$ / -87$^\circ$ & 1.9 & 9.8 \\
 HD~187885 & $0\farcs93 \times 0\farcs78$ / -77$^\circ$ & 1.9 & 8.0 \\ 
 Hen~3-1475 & $1\farcs22 \times 0\farcs65$ / -84$^\circ$& 1.6 & 10.0 \\
 M1-92 & $1\farcs10 \times 0\farcs77$ / -11$^\circ$ & 2.2 & 10.3 \\
\end{tabular}
\tablefoot{The beam FWHM is an average value for the spectral window with lowest frequencies
(between $\sim 219.44$~GHz and $\sim 221.31$~GHz) and the noise level in the spectral cube
is given at a reference frequency of 221~GHz towards the position of the continuum peak.}
\end{table}

\begin{table*}
\centering
\caption{Parameters obtained from the literature and measured values of the continuum emission for the sample stars.}
\label{tab:cont}
\begin{tabular}{ c @{\phantom{a}}c @{\phantom{a}}c @{\phantom{a}}c @{\phantom{}}c @{\phantom{}}c @{\phantom{}}c @{\phantom{a}}c
@{\phantom{~}}c @{\phantom{aa}}c @{\phantom{aa}}c }
Source & $d$ & $L$ & Chem. & Spec. & $\upsilon_{\rm LSR}$ & $S^{\rm cont}_\nu$ & $\alpha^{\rm cont}$ & $\delta^{\rm cont}$ & FWHM and PA of cont\\
 & [kpc] & [$10^3 L_\odot$] & Type & Type & [km~s$^{-1}$] & [mJy] & & & [$\arcsec \times~ \arcsec / ^\circ$] \\
\hline
GLMP~950 &  3.9$^{\rm a}$ & 6 & C$^{\rm c}$ & ? & 22 & $5.2 \pm 0.6$ & $\ra{19}{47}{24}{43}$ & $\dec{+29}{28}{13}{6}$ & $0.48 (0.04) \times 0.39 (0.03)/170 (20)$
\\
GLMP~953 & 3.8$^{\rm a}$ & 6 & C$^{\rm c}$ & ? & 42 & $119 \pm 12$ &  $\ra{19}{50}{08}{20}$ & $\dec{+25}{11}{57}{1}$
& $0.24(0.02) \times 0.26(0.01)/170(70)$\\
AFGL~5385 & 3.4$^{\rm a}$ & 12 & C$^{\rm d}$ & F4I$^{\rm f}$ & 7 & $12.9 \pm 1.4$ & $\ra{17}{47}{13}{50}$ & $\dec{-24}{12}{51}{6}$ &  $1.4 (0.1) \times 0.79(0.09)/59(8)$ \\
HD~187885 & 2.4$^{\rm a}$ & 6 & C$^{\rm e}$ & F0Ie$^{\rm f}$ & 25 & $6.3 \pm 0.7$ & $\ra{19}{52}{52}{70}$ & $\dec{-17}{01}{50}{3}$ &
$1.67(0.06)\times1.43(0.06)/39(12)$ \\
Hen~3-1475 & 5.8$^{\rm b}$ & 11 & O\phantom{$^{\rm e}$} & B7e$^{\rm b}$ & 49 & $29 \pm 3$ & $\ra{17}{45}{14}{18}$ & $\dec{-17}{56}{46}{9}$ &
$1.10(0.09) \times1.00(0.07)/ $-- \\
M1-92 & 2.6$^{\rm a}$ & 6 & O\phantom{$^{\rm e}$} & B0.5IV$^{\rm g}$ & 2 & $127 \pm 13$ & $\ra{19}{36}{18}{88}$ & $\dec{+29}{32}{49}{6}$ & -- / -- \\
\end{tabular}
\tablefoot{Adopted distances and corresponding luminosities, chemical types, spectral types, systemic velocities measured from the spectra,
and measured integrated continuum flux densities at 1.31~mm ($S^{\rm cont}_\nu$), continuum-peak positions, and FWHM sizes and position angle
from Gaussian fit to continuum images.}
\tablebib{(a) \citet{Vickers2015}; (b) \citet{Riera1995}; (c) \citet{Omont1993}; (d) \citet{Su1998}; (e) \citet{vanWinckel2000}; (f) \citet{Suarez2006}; (g) \citet{Allen1976}.}
\end{table*}

\section{Observations}
\label{sec:obs}

The observations were carried out using the Atacama Large Millimeter / submillimeter Array (ALMA) in the context of project 2021.1.01259.S.
The data reduction was performed by the observatory using the standard ALMA pipeline and
the Common Astronomy Software Applications for Radio Astronomy \citep[CASA,][]{CASA2022}
versions 6.2.1.7 for HD~187885 and 6.4.1.12 for the other five sources. We obtained the reduced cubes made available through
the ALMA archive.
The telescope setting was planned in Band~6 so that the \jtwoone\, lines of C$^{17}$O (at 224.714~GHz)
and C$^{18}$O (at 219.560~GHz) could be observed in the same frequency tuning.
The same line of $^{13}$CO (at 220.399 GHz) was also observed in the spectral window used to observe the C$^{18}$O line.
No $^{12}$CO line was observed as part of this program.

The four spectral windows covered frequencies ranging in the rest frame of the sources from $\sim 219.44$~GHz to $\sim 221.31$~GHz,
$\sim 222.93$~GHz to $\sim 224.80$~GHz, $\sim 234.05$~GHz to $\sim 234.92$~GHz, and $\sim 235.95$~GHz to $\sim 237.81$~GHz.
The exact frequency ranges varied slightly from one source to the other. We adopt an absolute flux calibration uncertainty
of $10\%$\footnote{ALMA Technical Handbook for Cycle 8}.
This does not apply to the calculations of line ratios, for which the relative flux calibration is much more accurate. Hence,
the relative flux calibration between the different molecular lines is negligible compared to the uncertainty from integrating the line emission.

The details of the obtained cubes are given in Table~\ref{tab:obs}.
The resolution of the maps varies between $0\farcs63$ and $0\farcs78$ for the different sources,
while the sensitivity in the continuum-subtracted maps varies between 1.6 and 2.5 mJy/beam for a channel
resolution of $\sim1$~MHz ($\sim 1.4$~km~s$^{-1}$).
The maximum recoverable scale (MRS) is significantly larger than the extent of all sources in any given channel, with the exception
of the continuum emission in M1-92 (see below). Even in this case, the observations presented here seem to recover all
continuum emission as discussed in Section~\ref{sec:M1-92}.

The continuum maps produced by the ALMA pipeline
include the four spectral windows and have a representative frequency 
of 228.6~GHz, which corresponds to a wavelength of 1.31~mm.

\section{Observational results}
\label{sec:obsRes}

The \jtwoone\, lines of the three targeted isotopologues of CO ($^{13}$CO, C$^{17}$O, and C$^{18}$O) were detected for all sources.
About $30$ other lines emitted by $\sim 10$ other molecules were also detected (Table~\ref{tab:lines}).
The vast majority of these lines was detected only towards the C-rich sources GLMP~950 and GLMP~953.
The detections towards these two sources include lines assigned by us to NaCl, SiC$_2$,
K$^{37}$Cl (tentative), and HC$_3$N. We also detect a few unidentified lines.
The spectra of the two other C-rich sources display almost no lines besides those from CO isotopologues, with only
an HC$_3$N line detected towards AFGL~5385, and no other lines detected towards HD~187885. The lack of lines
in the spectrum of HD~187885 might be caused by its low initial metallicity.

The two O-rich sources show a smaller number of lines. Hen~3-1475 is the most line rich of the two, with
five lines detected besides those from CO isotopologues. We assign the detected lines to
emission from SO, SiS, and SO$_2$. M1-92 shows emission lines of SO and SO$_2$.

\subsection{Methodology for line and continuum measurements}

To determine the sizes of the apertures for measuring the $^{13}$CO, C$^{17}$O, and C$^{18}$O~\jtwoone\, line emission and the continuum fluxes, we employ the following
procedure. First, we obtain the line profiles
from circular regions with 5\arcsec\, in diameter for each source. Then, we measure the full-width at zero intensity of the C$^{17}$O and C$^{18}$O~\jtwoone\, lines, which
are very similar for all sources.
Using the velocity range measured for the C$^{18}$O~\jtwoone\, line, we produced moment-zero (integrated in velocity) maps of the $^{13}$CO~\jtwoone\, line.
This was done to obtain a map of the brightest line ($^{13}$CO~\jtwoone\,) while excluding any velocity components not detected in C$^{18}$O~\jtwoone.
If the contour marking emission three times above the root mean square (3-$\sigma$) of this $^{13}$CO\,\jtwoone\, moment-zero map was larger than the 5\arcsec~ aperture, we used
an aperture of 7\arcsec\, instead, which proved sufficiently large to enclose the 3-$\sigma$ contour for all sources.
We use circular apertures of similar size when possible for simplicity because the uncertainties introduced by the aperture shape are small for 
sources that are nearly circular. Following this procedure, we obtain aperture diameters of
5\arcsec~ for GLMP~950 and Hen~3-1475 and
7\arcsec~ for GLMP~953, AFGL~5385 and HD~187885. For M1-92, we integrated the emission with a non-circular aperture equal to the 3-$\sigma$ contour.
This was motivated by the very elongated morphology and large size of the emission region in M1-92 (see Fig.~\ref{fig:13CO}).
The high-velocity emission discussed below observed towards different sources
are produced in regions within the circular apertures described above for all sources but Hen~3-1475.
Hence, to obtain the spectra of the high-velocity component of Hen~3-1475, we used a rectangular aperture oriented along the axis of the high-velocity emission as described below.
The obtained moment-zero maps of the C$^{13}$O, C$^{17}$O and C$^{18}$O~\jtwoone\, lines are shown in Figs.~\ref{fig:13CO}, \ref{fig:C17O}, and \ref{fig:C18O}, respectively.

The continuum emission is expected to be dominated by dust emission at these wavelengths,
with only a very small contribution from free-free emission at 1.3~mm for the two objects with B-type central stars,
Hen~3-1475 and M1-92 \citep[see e.g.,][]{SanchezContreras2017}.
The continuum flux density reported in Table~\ref{tab:obs}
were obtained by integrating the emission within the same apertures used to extract the line profiles, and, hence,
adapted to fully enclose the emitting area of each source.
The FWHM sizes of the continuum emission reported below were measured
using the Interactive 2D fitting tool in CASA to fit a Gaussian function convolved with the beam to the maps in the
image plane. Hence, the sizes we report are beam-deconvolved FWHM of the best-fitting Gaussian distributions.
The measured continuum flux density, position of the emission centre, and continuum
FWHM are given in Table~\ref{tab:cont}. All coordinates throughout the paper are given in the equatorial system (J2000).

\subsection{GLMP 950 (IRAS~19454+2920)}
\label{sec:GLMP950}

\begin{figure}[t]
\begin{center}
\includegraphics[width=0.30\textwidth,clip, trim = 10 5 25 0]{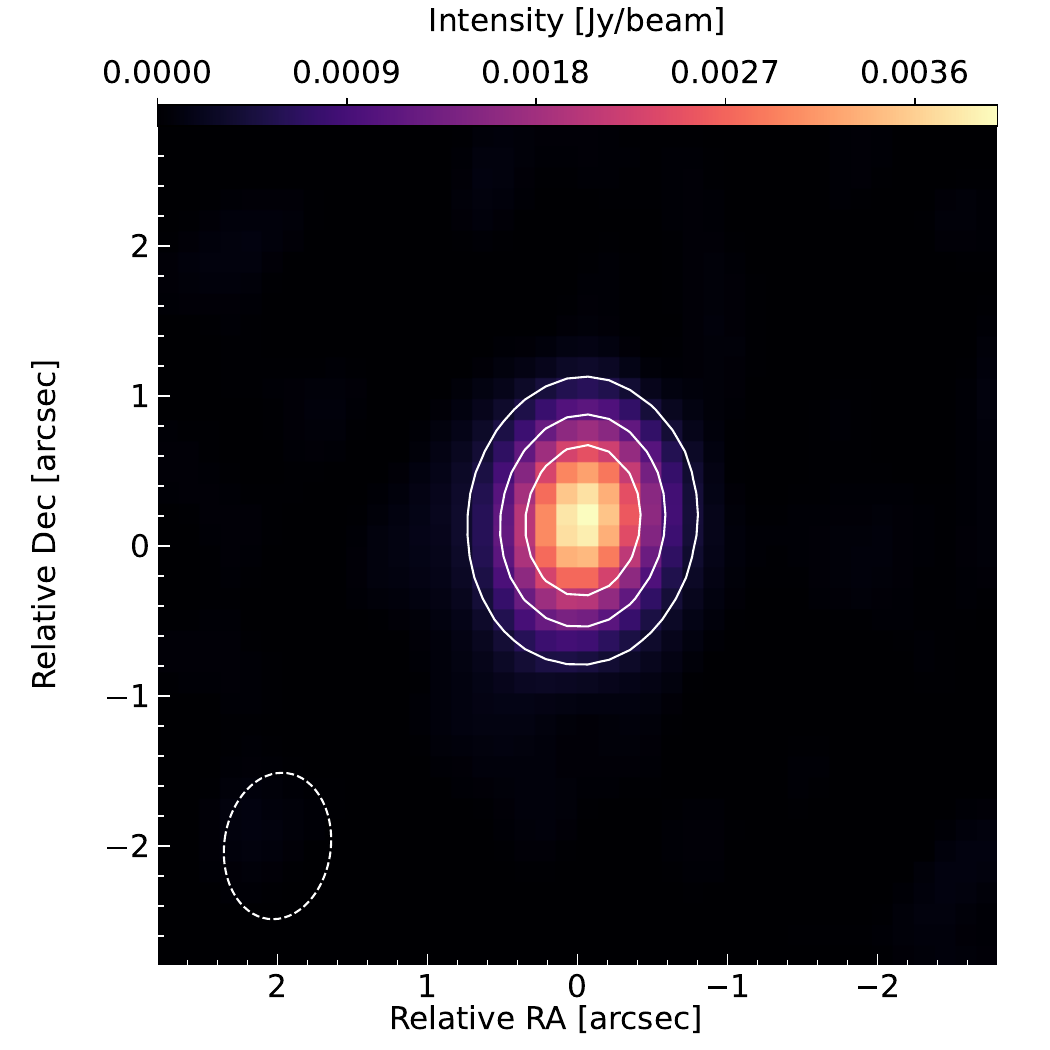}
\includegraphics[width=0.31\textwidth,clip, trim = 20 10 2 50]{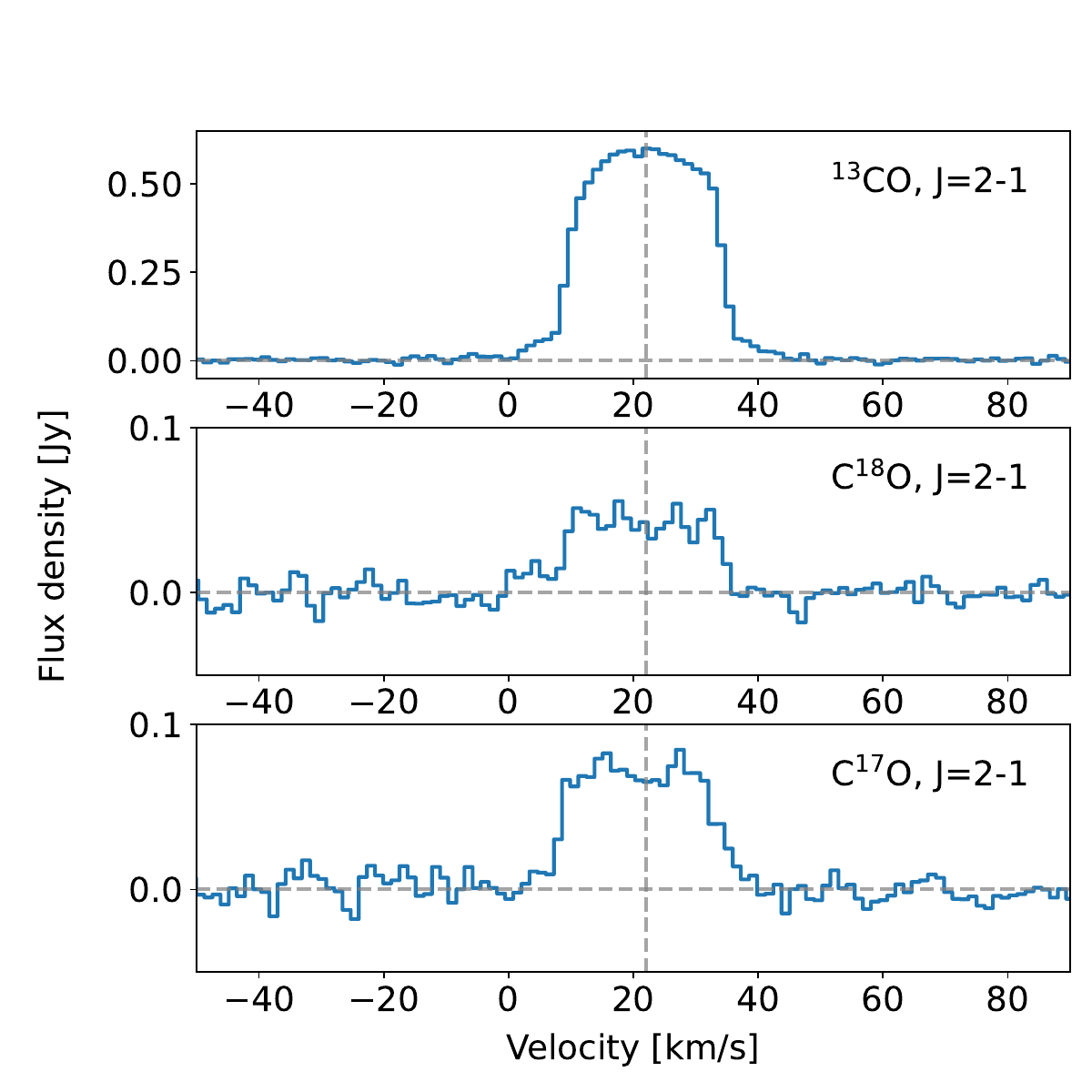}
 \caption{Continuum (1.3~mm) and \jtwoone\, $^{13}$CO, C$^{18}$O, and C$^{17}$O line emission towards GLMP~950.
 {\it Top:} Maps of the continuum (colour map) and $^{13}$CO ~\jtwoone\, emission integrated over the whole line emission (contours) at
 10\%, 30\%, 50\% and 70\% of the peak value (6.2~Jy~beam$^{-1}$~km~s$^{-1}$). The continuum emission peaks at
 at $\alpha = \ra{19}{47}{24}{430}$ and $\delta =\dec{29}{28}{13}{556}$.
 {\it Bottom:} source-integrated spectra of the $^{13}$CO~\jtwoone, C$^{18}$O~\jtwoone, and C$^{17}$O~\jtwoone\, lines.
 The vertical line marks the inferred velocity of the source $\upsilon_{\rm LSR} = 22$~km~s$^{-1}$.}
    \label{fig:GLMP950}
\end{center}
\end{figure}

The continuum image and the line profiles of the $^{13}$CO, C$^{17}$O, and C$^{18}$O~\jtwoone~
extracted towards GLMP~950 from a circular region centered on the source and with $5\arcsec$ in diameter are shown in Fig.~\ref{fig:GLMP950}.
The line fluxes obtained from integrating the \jtwoone\, lines from $^{13}$CO, C$^{17}$O, and C$^{18}$O over the same region 
are given in Table~\ref{tab:meas}.

\subsubsection{Continuum}

The original pointing direction was $\alpha = \ra{19}{47}{24}{835}$ and
$\delta = \dec{29}{28}{10}{817}$ based on the visible position assigned by the Simbad database \citep{Simbad2000}
at the time of our observations, which was that of the source Gaia DR3 2031794791233840128 \citep{Gaia2021}.
That position also corresponds to the near-infrared source 2MASS J19472480+2928108. However, we detected a mm continuum source at
$\alpha = \ra{19}{47}{24}{43}$ and $\delta =\dec{29}{28}{13}{556}$, which 
lies $5\farcs96$ away from the 2MASS and Gaia coordinates.
We inspected publicly available Spitzer/IRAC images downloaded from the IRSA archive, and noticed two individual sources, one at the optical/near-infrared position and another one at the location of our mm source (Fig.~\ref{fig:Spitzer_GLMP950}). The latter is brighter at longer Spitzer wavelengths. We conclude that our mm source corresponds to the far-infrared source IRAS~19454+2920 and the GLMP entry. Therefore, the position we report is the actual position of GLMP~950.

The fit of an elliptical Gaussian source to our continuum image reveals a deconvolved FWHM significantly smaller than the beam, with
a major axis of $485 \pm 40$~milliarcsecond (mas), a minor axis of $390 \pm 30$~mas
and a position angle of $170 \pm 20^\circ$.
{ The orientation we find is similar to that of the beam and, hence, it is unclear whether this an observational effect or a real feature.
Moreover, the orientation reported by \cite{Lagadec2011} in the mid-IR ($\sim 100^\circ$) is significantly different from the one we find.
Higher-angular-resolution observations are necessary to confirm whether this difference is real given the small size of the source compared
to the resolution of the observations we report.}
A flux density of $5.0 \pm 0.5$~mJy is recovered from the fit.
A slightly larger flux density, of $5.2\pm0.6$~mJy is obtained from integrating the continuum image within a region of $5\arcsec$ in diameter.

\subsubsection{Line emission}

The bulk of the emission in the $^{13}$CO~\jtwoone\, line is seen at velocities, $|\upsilon - \upsilon_{\rm LSR}| \lesssim 16$~km~s$^{-1}$ from the systemic velocity with respect to the
local standard of rest (LSR),
$\upsilon_{\rm LSR} \approx 22$~km~s$^{-1}$,
while a weaker component extends to $|\upsilon - \upsilon_{\rm LSR}| \lesssim 20$~km~s$^{-1}$. Integrating the $^{13}$CO~\jtwoone\, line over the interval
16~km~s$^{-1} \leq |\upsilon - \upsilon_{\rm LSR}| \leq$~20~km~s$^{-1}$ produces maps that peak essentially at the same position as that produced by integrating the central 
region of the line, $|\upsilon - \upsilon_{\rm LSR}| \leq$~16~km~s$^{-1}$. Hence, no sign
of a more complex morphology of the line emission as a function of velocity is detected
at the resolution of the observations we report.

A Gaussian fit to the map produced integrating the $^{13}$CO~\jtwoone\, line over the whole profile reveals an essentially circular source with
deconvolved FWHM of
major and minor axes of $1\farcs13 \pm 0\farcs08$ and $1\farcs02 \pm 0\farcs07$, respectively, and an unconstrained
position angle. 

Neither the continuum nor the moment-zero maps of the $^{13}$CO~\jtwoone\, line reveal a significantly elongated source. This differs from
the 2\arcsec east-west source size reported by \cite{Lagadec2011} from observations of dust emission in the mid-IR.

\subsection{GLMP 953 (IRAS~19480+2504)}
\label{sec:GLMP953}

\begin{figure*}[th!]
\begin{center} 

\includegraphics[width=0.32\textwidth,clip, trim = 10 5 25 0]{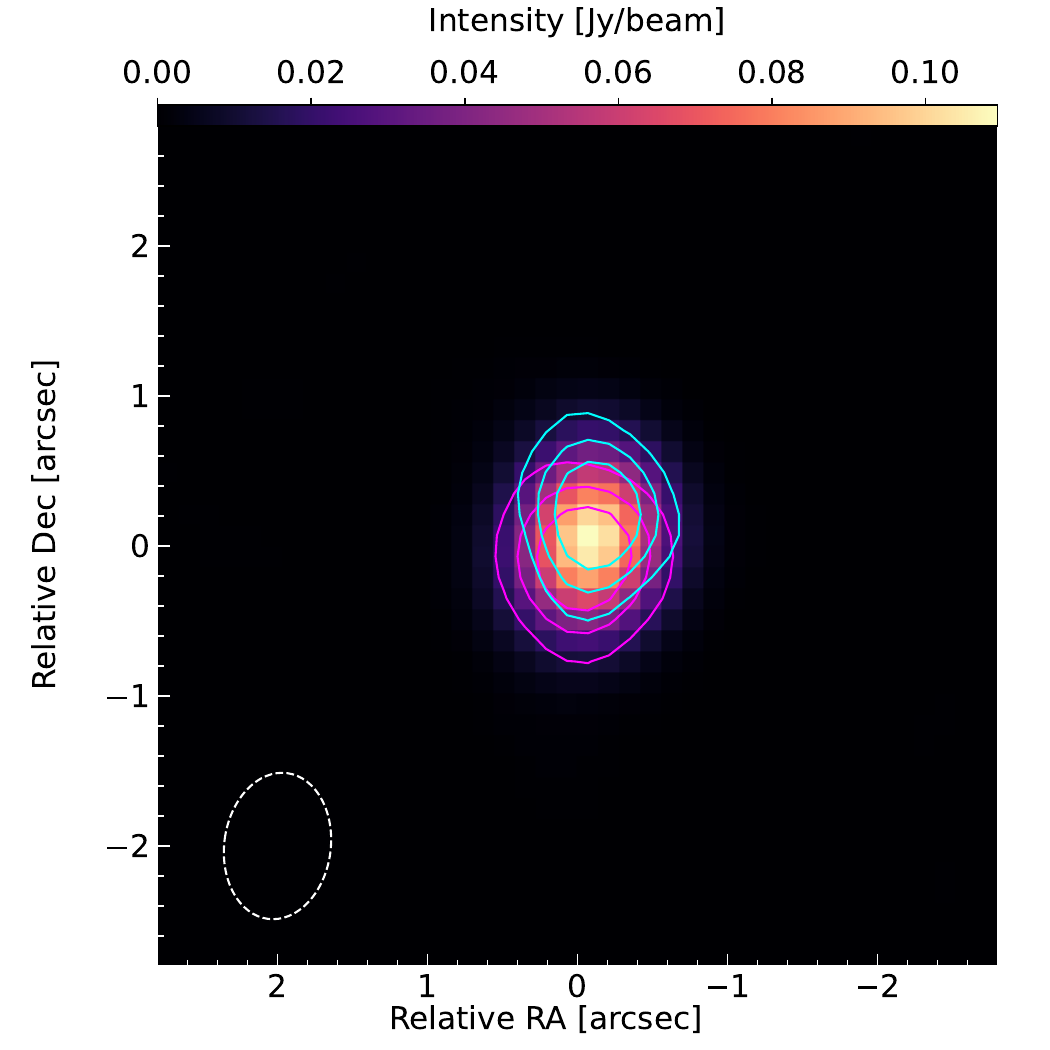}
\includegraphics[width=0.33\textwidth,clip, trim = 20 10 2 50]{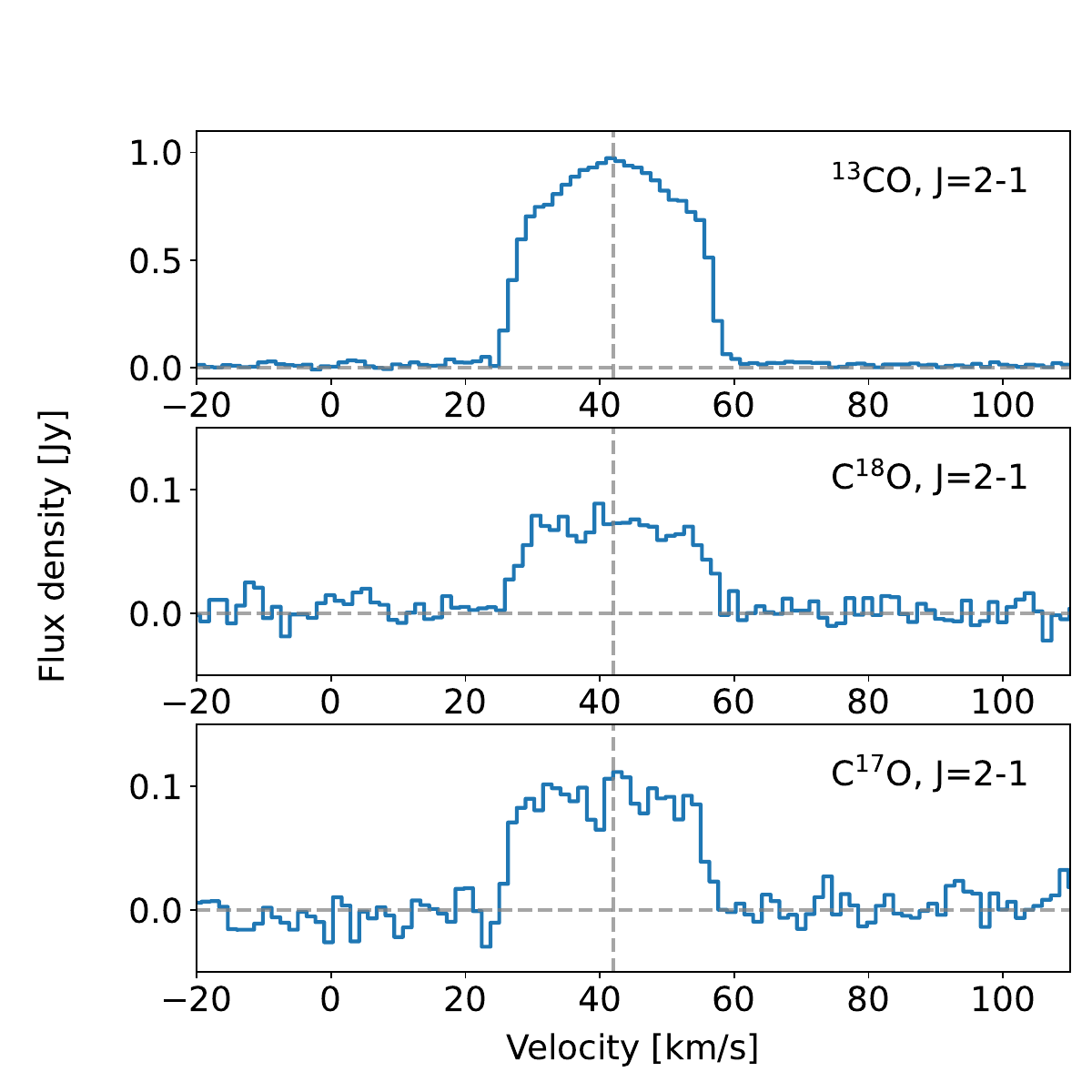}
\includegraphics[width=0.32\textwidth,clip, trim = 10 0 9 0]{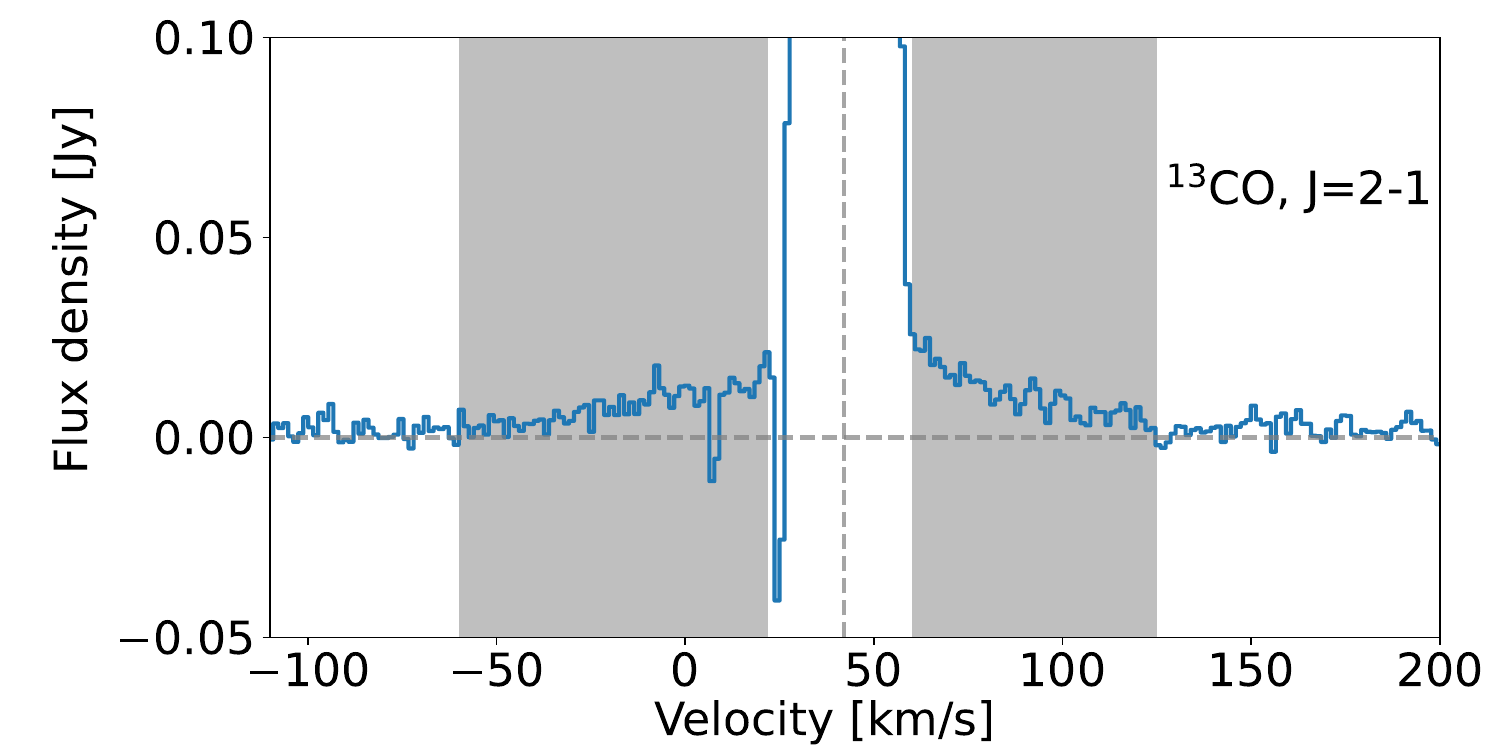}
 \caption{Continuum (1.3~mm) and \jtwoone\, $^{13}$CO, C$^{18}$O, and C$^{17}$O line emission towards GLMP~953.
 {\it Left:} Maps of the continuum (colour map) and high-velocity line emission (contours). The
 continuum peak is at $\alpha = \ra{19}{50}{08}{203}$ and $\delta = \dec{+25}{11}{57}{136}$.
 The full lines show
 emission in the $^{13}$CO~\jtwoone\, integrated over the spectral intervals marked in grey line plot. Magenta and cyan contours mark
 red-shifted and blue-shifted emission, respectively, at 10\%, 30\%, 50\%, and 70\% of the peak values of 0.009 (red-shifted)
 and 0.007 (blue-shifted) ~${\rm Jy~beam^{-1}\times km~s^{-1}}$. 
 {\it Middle:} spectra of the $^{13}$CO~\jtwoone, C$^{18}$O~\jtwoone, and C$^{17}$O~\jtwoone\, lines.
 {\it Right:} single-beam spectrum (to maximize the signal-to-noise ratio of the weak emission) 
 of the $^{13}$CO~\jtwoone\, line towards the continuum peak of GLMP~953 including a broader velocity range.}
   \label{fig:GLMP953}
\end{center}
\end{figure*}

The continuum image and the profiles of the $^{13}$CO, C$^{17}$O, and C$^{18}$O~\jtwoone\, lines extracted towards GLMP~953
using an aperture of 7\arcsec centered on the continuum peak are shown in Fig.~\ref{fig:GLMP953}.
The fluxes of the \jtwoone\, lines of $^{13}$CO, C$^{17}$O, and C$^{18}$O
integrated over the same region are given in Table~\ref{tab:meas}.

\subsubsection{Continuum}

Similarly to the case of GLMP 950 mentioned above, the original pointing direction was $\alpha = \ra{19}{50}{08}{281}$ and \mbox{$\delta = \dec{+25}{12}{00}{813}$},
based on the position reported in the Simbad database \citep{Simbad2000}, which corresponds to the optical source Gaia DR3 2026709201998745472 \citep{Gaia2021},
also positionally coincident with the near-IR source 2MASS J19500826+2512008 \citep{Cutri2003}. However, we detected a mm continuum source centered at
\mbox{$\alpha = \ra{19}{50}{08}{203}$} and $\delta = \dec{+25}{11}{57}{14}$. The position we find differs by 3\farcs83 from those in the Gaia and 2MASS catalogues.
Spitzer/IRAC images also show two distinct sources (Fig.~\ref{fig:Spitzer_GLMP953}), one coinciding with the position of the mm continuum and
another coinciding with the optical/near-IR source, which is significantly weaker in all Spitzer bands.
We thus conclude that our position (rather than the one reported by Simbad at the time of our observations) corresponds to that of the
GLMP entry and the far-IR source IRAS~19480+2504, and should be adopted as the actual position of GLMP~953.

By fitting an elliptical Gaussian to the mm-continuum source, we obtain a nearly circular intensity distribution, with deconvolved FWHM
of $240 \pm 20$~mas along the major axis and
$225 \pm 10$~mas along the minor axis,
which is roughly a factor of four smaller than the beam size.
The position angle and the flux density retrieved from the fit are $170 \pm 70^\circ$ and $118 \pm 12$ mJy, respectively.
Integrating the  continuum emission within an aperture of $5\arcsec$ in diameter reveals an essentially equal value of $119\pm12$~mJy.
The size we report is consistent with the unresolved source between 8 and 13~$\mu$m reported by \cite{Lagadec2011} based on observations
with a beam size of $\sim 0\farcs54$.

\subsubsection{Line emission}

Most of the emission in the $^{13}$CO~\jtwoone\, line lies within $|\upsilon - \upsilon_{\rm LSR}| \lesssim 16.5$~km~s$^{-1}$ with $\upsilon_{\rm LSR} \approx 42$~km~s$^{-1}$.
A weaker, higher-velocity component extending roughly between $-60$~km~s$^{-1}$ and 125~km~s$^{-1}$ is also observed. Hence, the total line width is $\sim 185$~km~s$^{-1}$.
Integrating the emission from the high-velocity component, from $|\upsilon - \upsilon_{\rm LSR}| \gtrsim 16.5$~km~s$^{-1}$ to the extreme of the line wings
(range marked in gray in Fig.~\ref{fig:GLMP953}), reveals emission regions slightly offset from the centre of the continuum peak. The distance between
the centers of the red- and blue-shifted high-velocity components is $\sim 0\farcs35$ and
the position angle is $\sim -10^\circ$, with a high associated uncertainty because of the small separation between
the two components.

\subsection{AFGL 5385 (IRAS~17441-2411, the Silkworm Nebula)}

\begin{figure}[th!]
\begin{center}
\includegraphics[width=0.32\textwidth,clip, trim = 10 5 25 0]{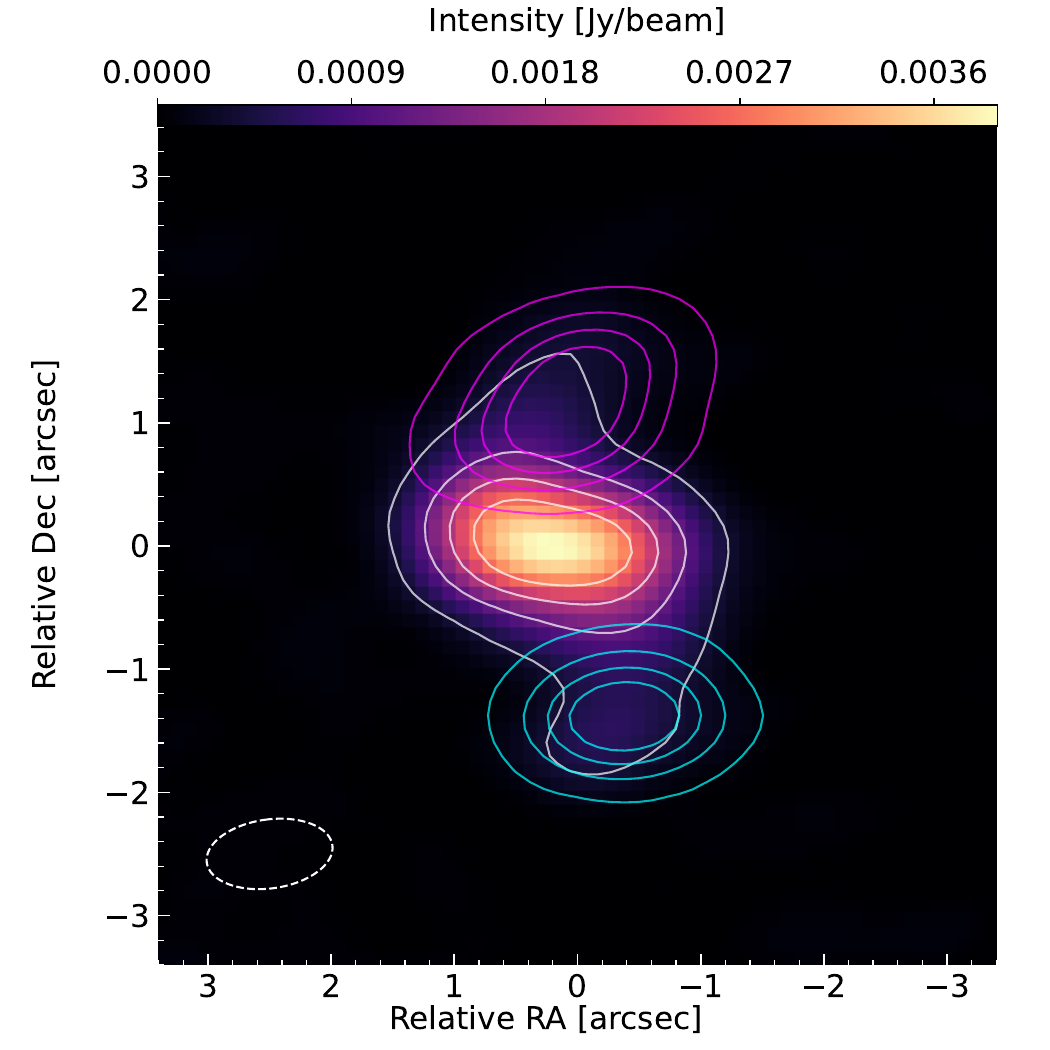}
\includegraphics[width=0.33\textwidth,clip, trim = 20 10 2 50]{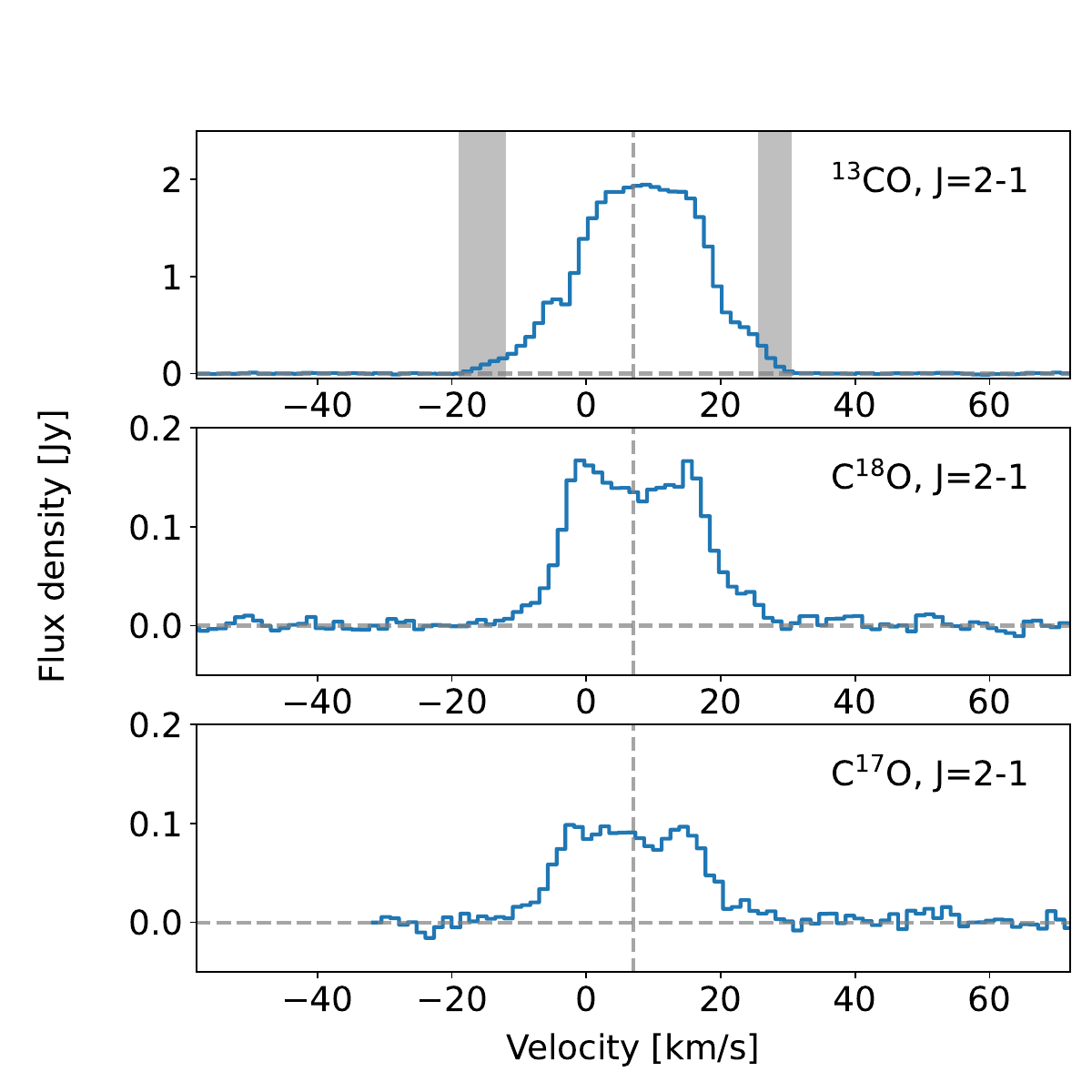}
 \caption{Continuum (1.3~mm) and \jtwoone\, $^{13}$CO, C$^{18}$O, and C$^{17}$O line emission towards AFGL~5385.
 {\it Top:} Maps of the continuum (colour map) and high-velocity line emission (contours). 
 The continuum peaks at $\alpha = \ra{17}{47}{13}{499}$ and $\delta = \dec{-24}{12}{51}{617}$.
 The full white line shows the continuum emission at 10\%, 30\%, 50\%, and 70\% of the peak value.
 The full colored lines show
 emission in the $^{13}$CO~\jtwoone\, integrated over the spectral intervals marked in grey in the line spectra.
 Magenta and cyan contours mark
 red-shifted and blue-shifted emission, respectively, at 10\%, 30\%, 50\%, and 70\% of the peak values of 0.068 (red-shifted)
 and 0.070 (blue-shifted) ~${\rm Jy~beam^{-1}\times km~s^{-1}}$.
 {\it Bottom:} source-integrated spectra of the $^{13}$CO~\jtwoone, C$^{18}$O~\jtwoone, and C$^{17}$O~\jtwoone\, lines.}
   \label{fig:AFGL_highVel}
\end{center}
\end{figure}

The continuum image and the profiles of the $^{13}$CO, C$^{17}$O, and C$^{18}$O~\jtwoone\, lines extracted towards AFGL~5385
using a circular region with $7\arcsec$ in diameter are shown in Fig.~\ref{fig:AFGL_highVel}.
The fluxes of the \jtwoone\, lines of $^{13}$CO, C$^{17}$O, and C$^{18}$O integrated over the same region are given in Table~\ref{tab:meas}.

\subsubsection{Continuum}

Fitting a Gaussian source, we obtain deconvolved FWHM axes of
$1\farcs4 \pm 0\farcs1$ and
$0\farcs79 \pm 0\farcs09$ and a position angle of $59^\circ \pm 8^\circ$. The integrated flux density obtained from
the fit is $11.8 \pm 1.3$~mJy.
The continuum map shows an S-shaped morphology with two extensions in a roughly
perpendicular direction to the elliptical central source. These reach $\sim 2\farcs3$ in radius from the centre,
and are not captured by the Gaussian fit. Including these extensions in the continuum integration using the $7\arcsec$ aperture
leads to a flux density of $12.9\pm1.4$~mJy. 

An additional point-like source is detected at \mbox{$\alpha = \ra{17}{47}{14}{03}$ and $\delta = \dec{-24}{12}{52}{8}$},
i.e. at $\sim 7\farcs5$ ($\sim 2.5\times10^4$~au at 3.4~kpc) to the east of AFGL~5385. The flux density of this second source
peaks at 0.4~mJy, which is $\sim 9$ times above the root-mean-square noise level of the continuum map. The continuum
image including this additional source is shown in Fig.~\ref{fig:AFGL_cont}. It is unclear whether this second 1.3~mm continuum source is associated to AFGL~5385,
or is a field object. { The brightness ratio between the two sources at 1.3~mm and the coincidence of the position of the 1.3~mm emission
of AFGL~5385 and that from the 2MASS maps imply that this secondary source does not affect the photometry obtained towards AFGL~5385 significantly.}

\subsubsection{Line emission}

We obtain $\upsilon_{\rm LSR}\approx 7$~km~s$^{-1}$ based on the approximate centres of the
C$^{18}$O and C$^{17}$O~\jtwoone\, lines.
This value, however, leads to the main component of the $^{13}$CO~\jtwoone\, line to show more red-shifted than blue-shifted emission,
which could be a sign of relatively high optical depths affecting the $^{13}$CO line shape as this produces the well-known effect of
blue-wing self-absorption \citep{Morris1985}.

The bulk of the $^{13}$CO~\jtwoone\, line emission arises
from $ -2.5~{\rm km}$~s$^{-1} \lesssim \upsilon \lesssim 20$~km~s$^{-1}$. A weaker, higher-velocity component extending from $\upsilon \approx -19$~km~s$^{-1}$
to $\upsilon \approx 30$~km~s$^{-1}$ is also present.
Interestingly, the line shape of this higher-velocity component is
much more symmetric about the systemic velocity. This reinforces the suggestion that the
velocity shift observed in the bulk of the $^{13}$CO~\jtwoone\, emission is due to optical depth effects in the CSE.

Integrating the higher-velocity components in frequency over the interval marked in gray in Fig.~\ref{fig:AFGL_highVel}
reveals emission produced by gas offset from the position of the continuum.
The displacement of the emission peak in the integrated images
with respect to the continuum peak is $\sim 1\farcs1$ for the red-shifted emission and $\sim 1\farcs5$ for the blue-shifted emission.
The high-velocity gas appears where the extensions of the continuum emission are also seen.

The position angle of the continuum emission region at millimeter wavelengths ($59^\circ \pm 8^\circ$)
{ differs significantly from that observed in the mid-IR of $\sim 130^\circ$ \citep{Volk2007,Lagadec2011}. Nonetheless,} the S-shaped profile seen in those images is also
clearly visible in the images we present. The emission regions of the
higher-velocity gas are aligned roughly in the north-south direction, in apparent agreement with the orientation of the bipolar nebula seen in
visible light \citep{Su1998}. This high-velocity emission also follows the tips of the S-shaped continuum emission region.
The total size of the emission region of the $^{13}$CO~\jtwoone\, line is $\sim 5\arcsec$, which is comparable to the size of the
bipolar lobes in the {\it HST} images \citep{Ueta2007} and of the mid-IR nebulosity \citep{Lagadec2011}.

\begin{figure*}[th!]
\begin{center}
\includegraphics[width=0.32\textwidth,clip, trim = 10 5 25 0]{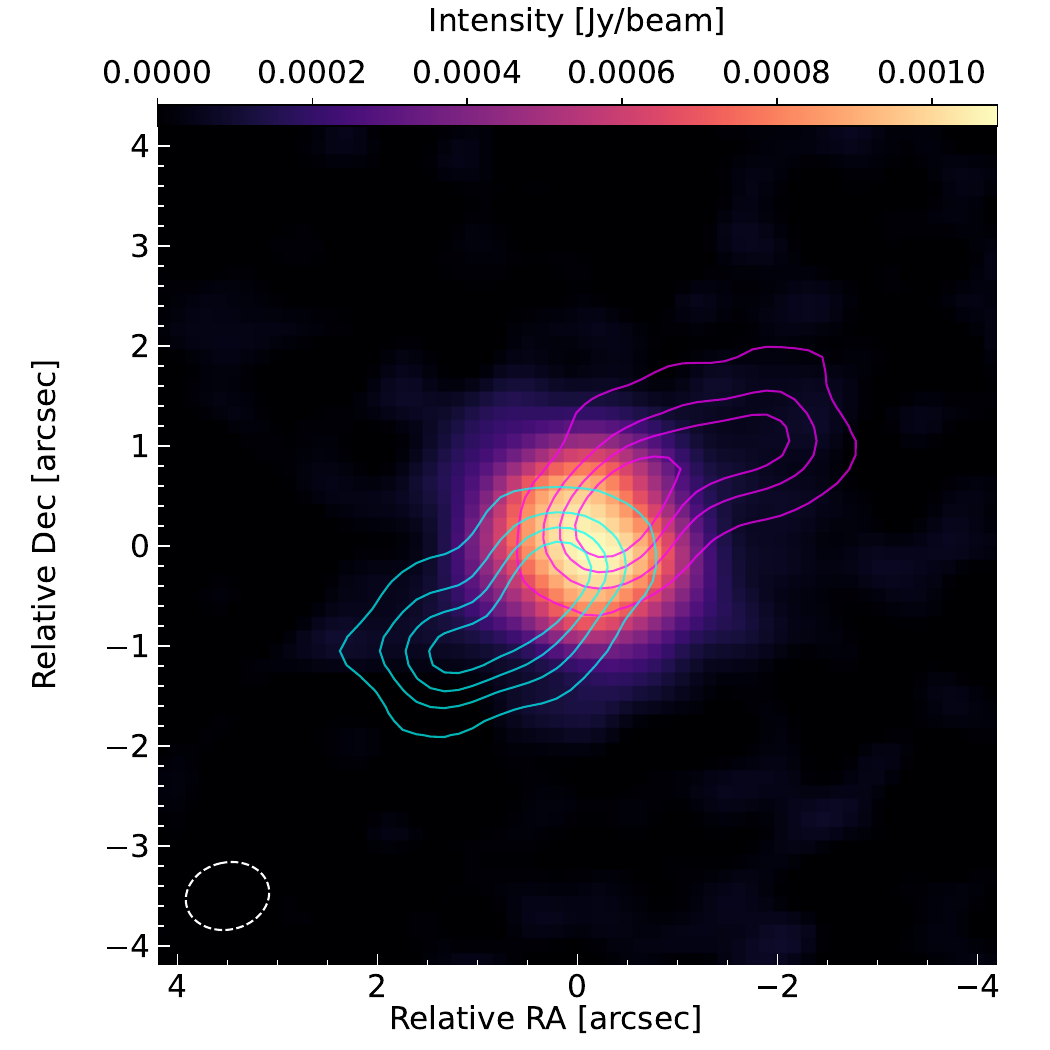}
\includegraphics[width=0.33\textwidth,clip, trim = 20 10 2 50]{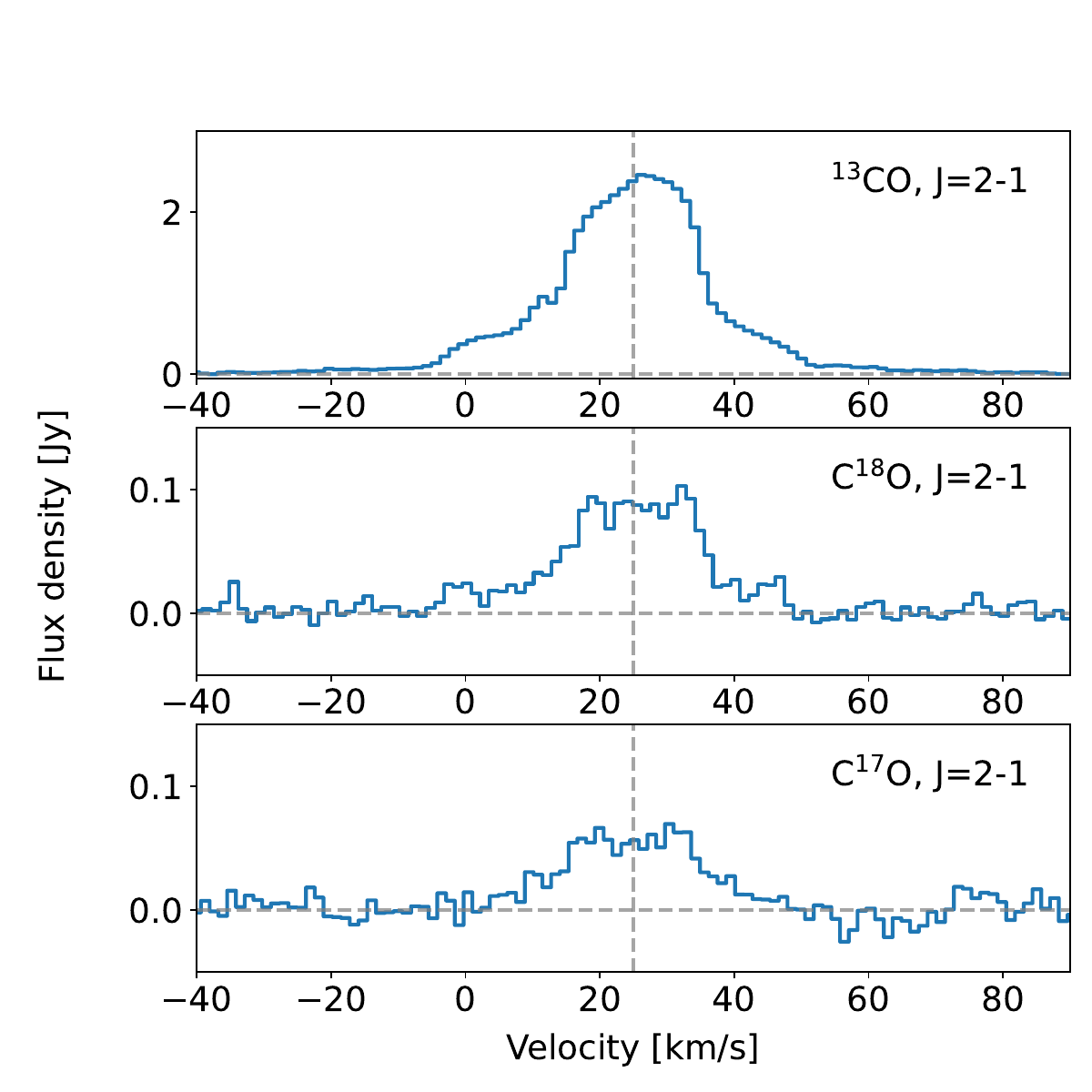}
\includegraphics[width=0.32\textwidth,clip, trim = 10 0 9 0]{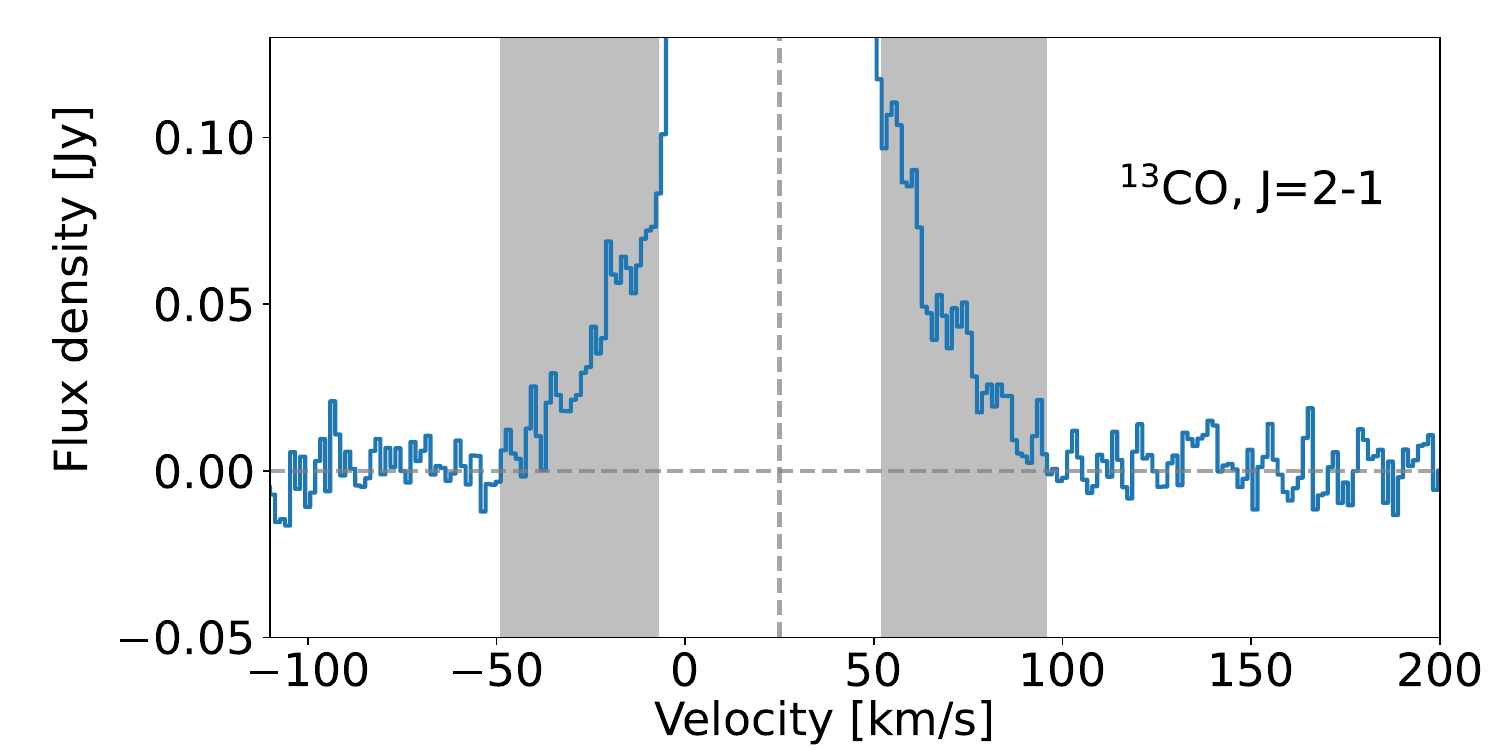}
 \caption{Continuum (1.3~mm) and \jtwoone\, $^{13}$CO, C$^{18}$O, and C$^{17}$O line emission towards HD~187885.
 {\it Left:} Maps of the continuum (colour map) and high-velocity line emission (contours).
 The continuum peaks at $\alpha = \ra{19}{52}{52}{691}$ and $\delta = \dec{-17}{01}{50}{500}$.
 The full lines show
 emission in the $^{13}$CO~\jtwoone\, integrated over the spectral intervals marked in grey line plot. Magenta and cyan contours mark
 red-shifted and blue-shifted emission, respectively, at 10\%, 30\%, 50\%, and 70\% of the peak values of 0.0160 (red-shifted)
 and 0.0146 (blue-shifted) ~${\rm Jy~beam^{-1}\times km~s^{-1}}$.
  {\it Middle:} source-integrated spectra of the $^{13}$CO~\jtwoone, C$^{18}$O~\jtwoone, and C$^{17}$O~\jtwoone\, lines.
 {\it Right:} source-integrated, zoomed-in spectrum of the $^{13}$CO~\jtwoone\, line showing the weak high-velocity emission.}
   \label{fig:HD187885_highVel}
\end{center}
\end{figure*}

\subsection{HD 187885 (IRAS~19500-1709)}

The continuum image and the profiles of the $^{13}$CO, C$^{17}$O, and C$^{18}$O~\jtwoone\, lines extracted towards HD~187885
using a circular region with 7\arcsec\, in diameter are shown in Fig.~\ref{fig:HD187885_highVel}.
The fluxes of the \jtwoone\, lines of $^{13}$CO, C$^{17}$O, and C$^{18}$O integrated over the same region are given in Table~\ref{tab:meas}.

\subsubsection{Continuum}

The continuum source is spatially resolved by the observations,
with a Gaussian fit revealing an elongated morphology with a deconvolved FWHM of $1\farcs67 \pm 0\farcs06$ for the major axis and 
of $1\farcs43 \pm 0\farcs06$ for the minor axis and a position angle of $39^\circ \pm 12^\circ$. The flux density obtained with the fit is $6.1 \pm 0.6$~mJy. 
Integrating the continuum using an aperture of 7\arcsec\, in diameter reveals a similar flux density of $6.3\pm0.3$~mJy.
This value is consistent with the upper limit reported for the flux density of this source at 1.2~mm of 23~mJy by \cite{Dehaes2007}.
{ The observations we report do not reveal an inner cavity in the disk as spotted in the mid-IR by \cite{Lagadec2011} and suggest based on modelling
by \cite{Clube2004} before, but this is most likely due to the worse spatial resolution of the ALMA data relatively to that from VISIR.}

\subsubsection{Line emission}

The profile of the $^{13}$CO~\jtwoone\, line (Fig.~\ref{fig:HD187885_highVel}) shows three different components. The central and intermediate
components comprise velocities $|\upsilon - \upsilon_{\rm LSR} | \lesssim 11$~km~s$^{-1}$ and $|\upsilon - \upsilon_{\rm LSR} | \lesssim 27$~km~s$^{-1}$, respectively, with $\upsilon_{\rm LSR} = 25$~km~s$^{-1}$.
The high-velocity component is significantly weaker and spans a velocity range of $\sim 145$~km~s$^{-1}$.

Integrating the emission of this high-velocity component over the spectral region indicated in gray
in Fig.~\ref{fig:HD187885_highVel} reveals two elongated structures
that extend away from the continuum peak in opposite directions. The elongations also curve in opposite directions, suggesting
a point-symmetric structure.
The continuum is elongated in a direction perpendicular to that of the high-velocity outflow. { The red-shifted feature seems to align well with an emission hole in the 8.59~$\mu$m images
presented by \cite{Lagadec2011}. Whether there is a physical connection between these two features is unknown.}

The peak value of the $^{13}$CO~\jtwoone\, line (2.45~Jy) is compatible with the value of 2.6~Jy observed by \cite{Bujarrabal2001}
using the IRAM~30m telescope. We converted the main beam temperature presented by the authors to flux density
using historic values for the efficiencies of the IRAM~30m telescope available at the website of the observatory.
Therefore, given the similarity between the two flux values and the beam size of the IRAM~30m telescope, which covers the entire extent of the nebula,
the structures traced by our ALMA observations contain most of the emission in the $^{13}$CO~\jtwoone\, line,
with no significant emission coming from more extended structures.

\begin{figure*}[th!]
\begin{center}
\includegraphics[width=0.32\textwidth,clip, trim = 10 5 20 0]{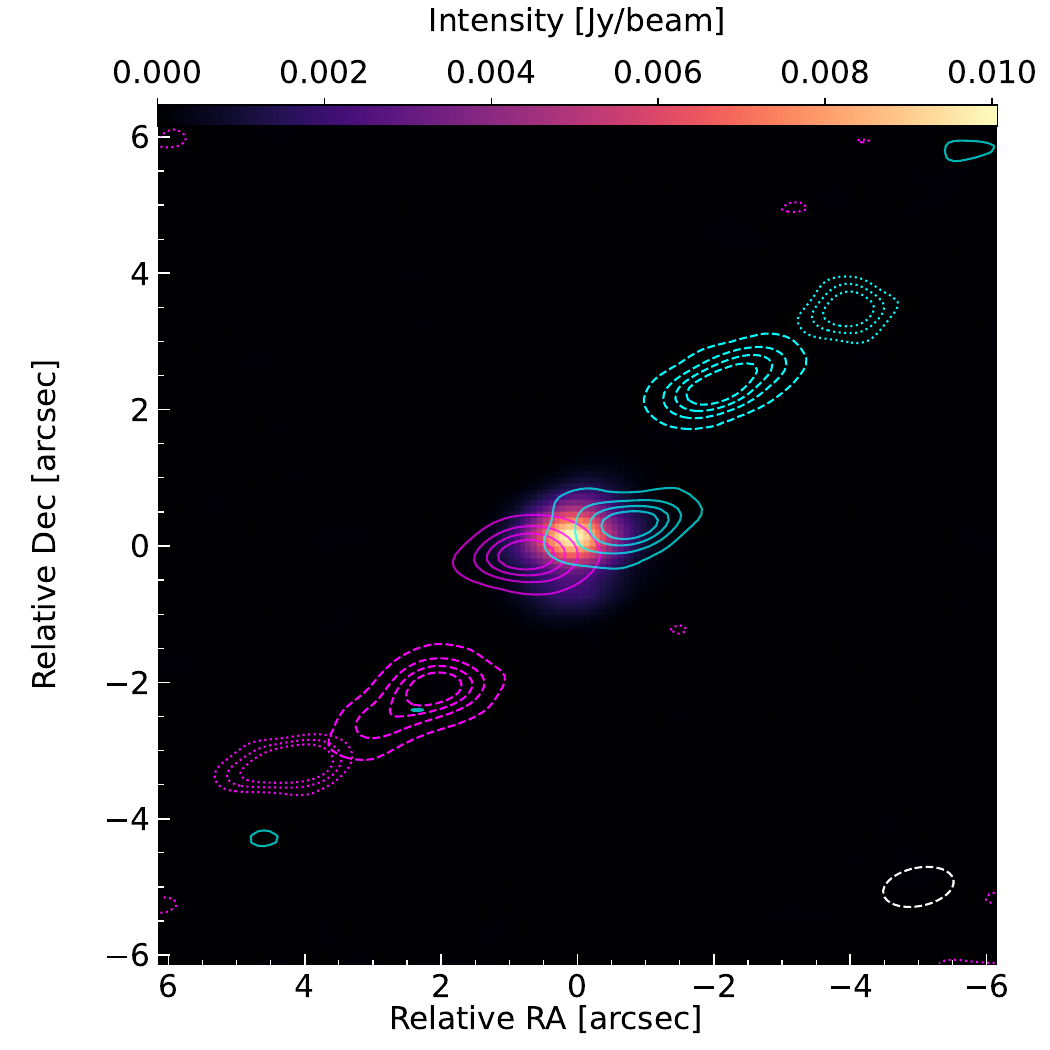}
\includegraphics[width=0.33\textwidth,clip, trim = 20 10 2 50]{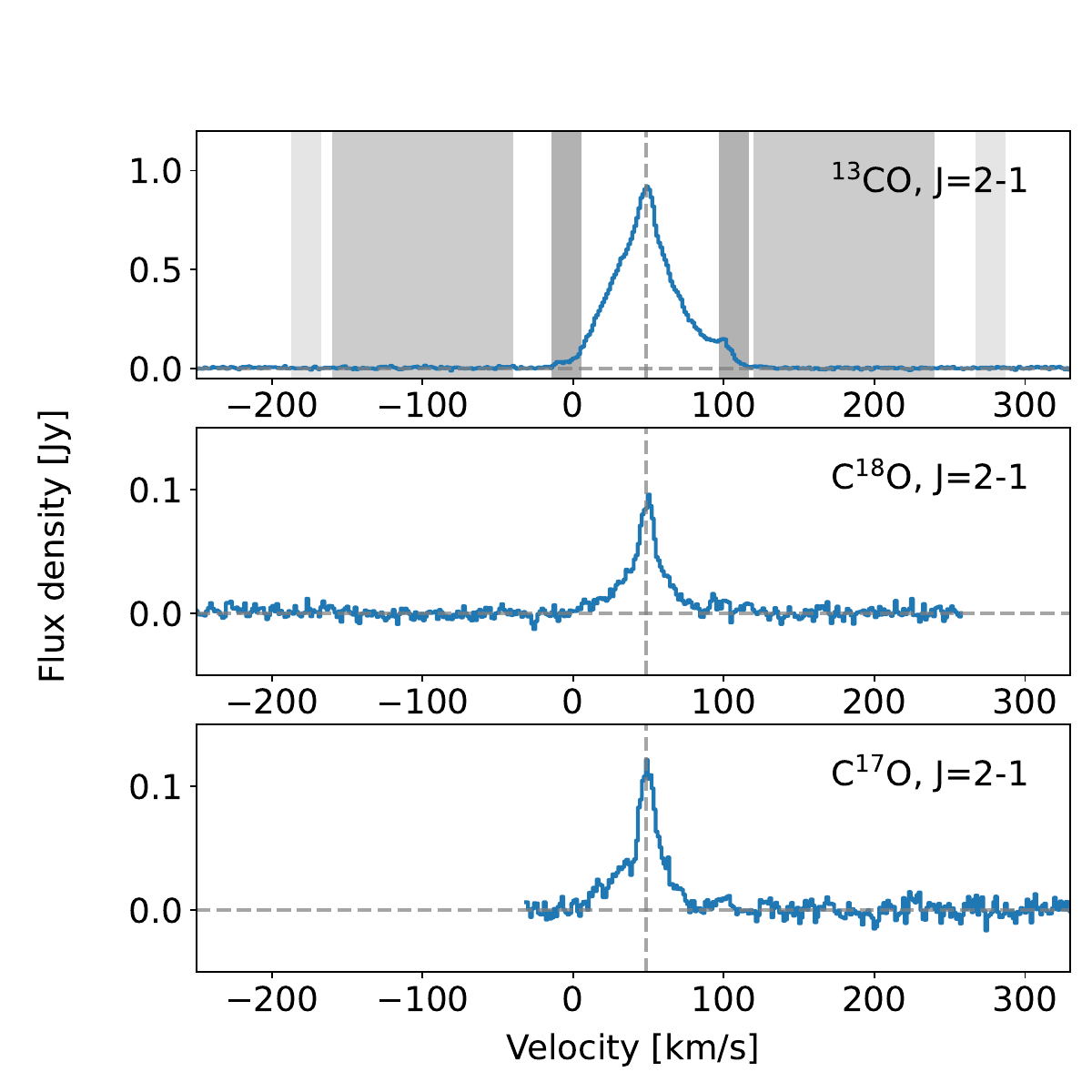}
\includegraphics[width=0.32\textwidth,clip, trim = 10 0 9 0]{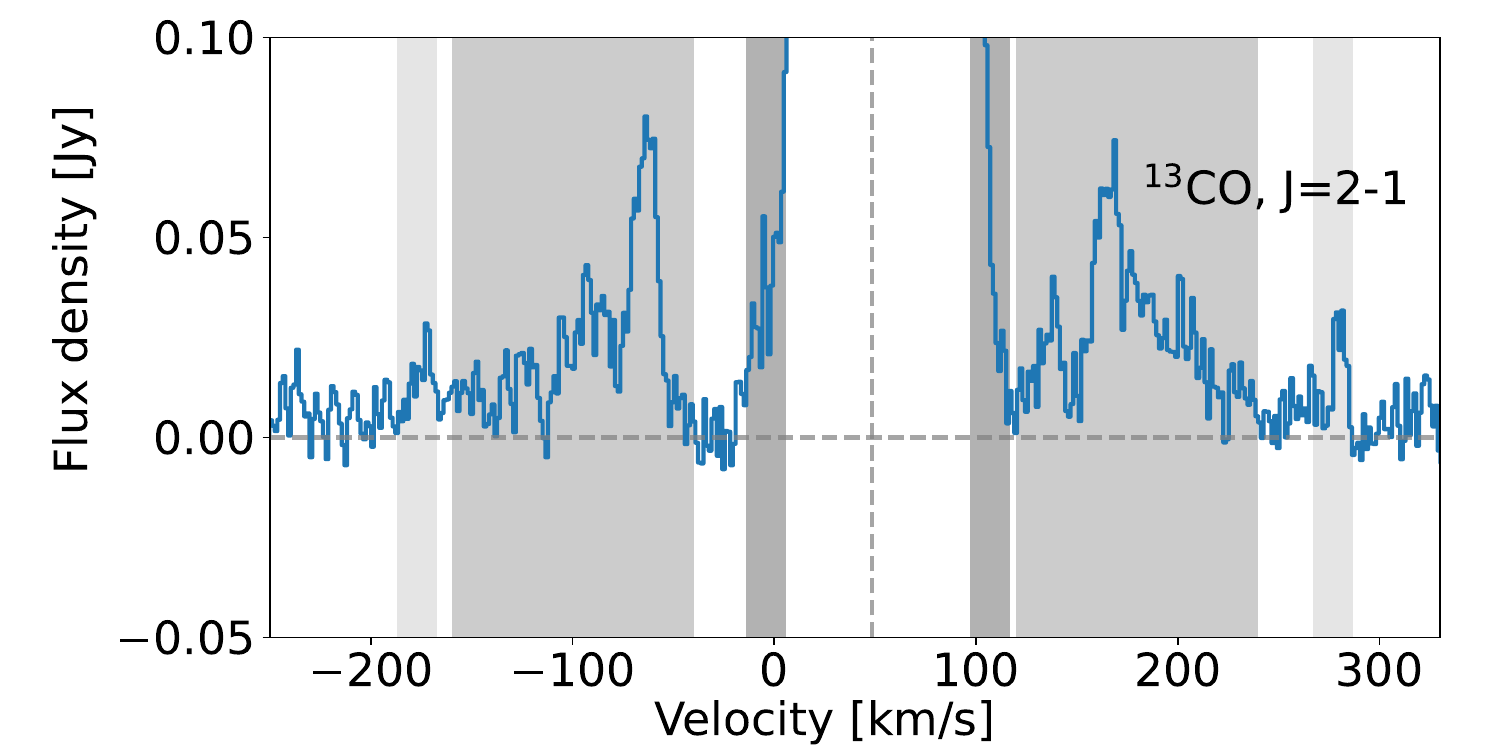}
 \caption{Continuum (1.3~mm) and \jtwoone\, $^{13}$CO, C$^{18}$O, and C$^{17}$O line emission towards Hen~3-1475.
 {\it Left:} Maps of the continuum (colour map) and high-velocity line emission (contours).
 The continuum peaks at $\alpha = \ra{17}{45}{14}{177}$ and $\delta = \dec{-17}{56}{46}{895}$.
 The full, dashed and dotted lines show
 emission in the $^{13}$CO~\jtwoone\, integrated over the spectral intervals at increasingly large expansion velocities
 marked in dark grey, grey, and light grey, respectively, in the
 line plots. Magenta and cyan contours mark
 red-shifted and blue-shifted emission, respectively, at 10\%, 30\%, 50\%, and 70\% of the peak value of each emission component.
 The 10\% value of the most extreme emission is not shown because it is comparable to the noise level.
 The peak values for the
 dark grey, grey, and light grey spectral regions are, respectively, 0.0422, 0.0122, and 0.0030~${\rm Jy~beam^{-1}\times km~s^{-1}}$ for the red-shifted emission and 0.0205, 0.0156, and 0.0083~${\rm Jy~beam^{-1}\times km~s^{-1}}$ for the blue-shifted emission.
 {\it Middle:} source-integrated spectra of the $^{13}$CO~\jtwoone, C$^{18}$O~\jtwoone, and C$^{17}$O~\jtwoone\, lines.
 {\it Right:} source-integrated spectrum of the $^{13}$CO~\jtwoone\, line showing the very-high-velocity components
 extracted using a rectangular aperture of $3\farcs5 \times 17\arcsec$ and oriented along the high-velocity-emission direction.}
   \label{fig:Hen_highVel}
\end{center}
\vspace{-0.2cm}
\end{figure*}

\subsection{Hen 3-1475 (IRAS~17423-1755)}

The continuum image and the profiles of the $^{13}$CO, C$^{17}$O, and C$^{18}$O~\jtwoone\, lines extracted towards Hen~3-1475
using a circular region with 5\arcsec\, in diameter are shown in Fig.~\ref{fig:Hen_highVel}.
The fluxes of the \jtwoone\, lines of $^{13}$CO, C$^{17}$O, and C$^{18}$O integrated over the same region are given in Table~\ref{tab:meas}.

\subsubsection{Continuum}

The continuum emission is spatially resolved in the observations and a fit of an elliptical Gaussian source provides a deconvolved FWHM of
$1\farcs10 (\pm 0\farcs09) \times 1\farcs00 (\pm 0\farcs07)$. Since the source is closer to appearing circular on the sky,
the position angle is not well constrained. The Gaussian fit reveals a flux density of $27 \pm 3$~mJy. By integrating the continuum using a
circular aperture of $5\arcsec$ in diameter, we obtain a flux density of $29\pm3$~mJy. These values are consistent with the previous
determination of $31 \pm 4$~mJy \citep{Huggins2004}.

\subsubsection{Line emission}

The $\upsilon_{\rm LSR}$ of $\sim 49$~km~s$^{-1}$
was derived using the peak of the narrow feature in the C$^{18}$O and C$^{17}$O lines, and it agrees fairly well with the
peak of the $^{13}$CO line.

Different velocity components are observed in the lines of the CO isotopologues.
A central core is apparent in the lines of C$^{18}$O and C$^{17}$O between velocities
$|\upsilon - \upsilon_{\rm LSR}| \lesssim 9$~km~s$^{-1}$ (with $\upsilon_{\rm LSR} = 48.5$~km~s$^{-1}$), but is less clearly discernible in the $^{13}$CO line.
Wings extending up to $|\upsilon - \upsilon_{\rm LSR}| \lesssim 45$~km~s$^{-1}$
are prominent even in the lines of the two rarer isotopologues. In the $^{13}$CO line, the wings extend to even higher velocities
$|\upsilon - \upsilon_{\rm LSR}| \lesssim 65$~km~s$^{-1}$.
The difference between the velocity range spanned by the wings in the different lines is most likely only due to the higher signal-to-noise
in the $^{13}$CO line.
The tips of these wings are marked in dark gray in Fig.~\ref{fig:Hen_highVel}.
Integrating the line emission over this spectral region reveals
emission regions with peaks offset from the continuum peak emission
by $\sim 1\farcs0$ and $\sim 1\farcs1$ for the blue-shifted and red-shifted components, respectively.

The peak of the emission in the $^{13}$CO~\jtwoone\, line ($\sim 0.9$~Jy) is comparable to the single-dish measurements
of $\sim 0.7$~Jy \citep{Bujarrabal2001}. The flux density was again
calculated using historic values for the efficiencies of the IRAM~30m telescope. The difference
is most likely due to the uncertainty in flux calibration of the single-dish data and
implies that no flux is lost in the ALMA data we present.

At larger scales (up to $\sim 8\arcsec$ from the central source), we find an even weaker component of the $^{13}$CO line, which is
marked in intermediate-dark and light gray in Fig.~\ref{fig:Hen_highVel}. To maximize the signal-to-noise
of the high-velocity spectral features, we extracted spectra using a rectangular aperture of $3.5\arcsec \times 17\arcsec$ and oriented along the high-velocity-emission direction
(Fig.~\ref{fig:Hen_highVel}).
This emission appears at more extreme projected velocities up to $\sim 240$~km~s$^{-1}$ from the $\upsilon_{\rm LSR}$.
Integrating over the intermediate-dark spectral regions
reveals two emission regions with peak emission offset from the continuum peak by
$\sim 4\arcsec$ for the blue- and red-shifted components. Emission at the extreme-velocity component (marked in light gray),
at projected velocities $\sim 240$~km~s$^{-1}$ from the $\upsilon_{\rm LSR}$, are produced still farther away. Hence, there is a positive velocity gradient in
this high-velocity, point-symmetric molecular emission.
To our knowledge, this is the first time these components with projected expansion velocities $> 50$~km/s have been detected towards Hen~3-1475 in observations of molecular
lines. Previous observations recovered only the central component of the line with projected
velocities up to $\sim 50$~km~s$^{-1}$ from the $\upsilon_{\rm LSR}$ \citep[e.g.][]{Bujarrabal2001,Huggins2004}.

Interestingly, two different orientation directions can be identified in the high-velocity outflow. The inner, lower-velocity emission component displays a position angle
relative to the north direction of $\sim -70^\circ$, while the outer, higher-velocity components display a position angle of $\sim -45^\circ$. In both cases,
point symmetry about the continuum peak is observed.

\subsection{M1-92 (IRAS 19343+2926, Minkowski's Footprint)}
\label{sec:M1-92}
\begin{figure}[th]
\begin{center}
\includegraphics[width=0.32\textwidth,clip, trim = 10 5 25 0]{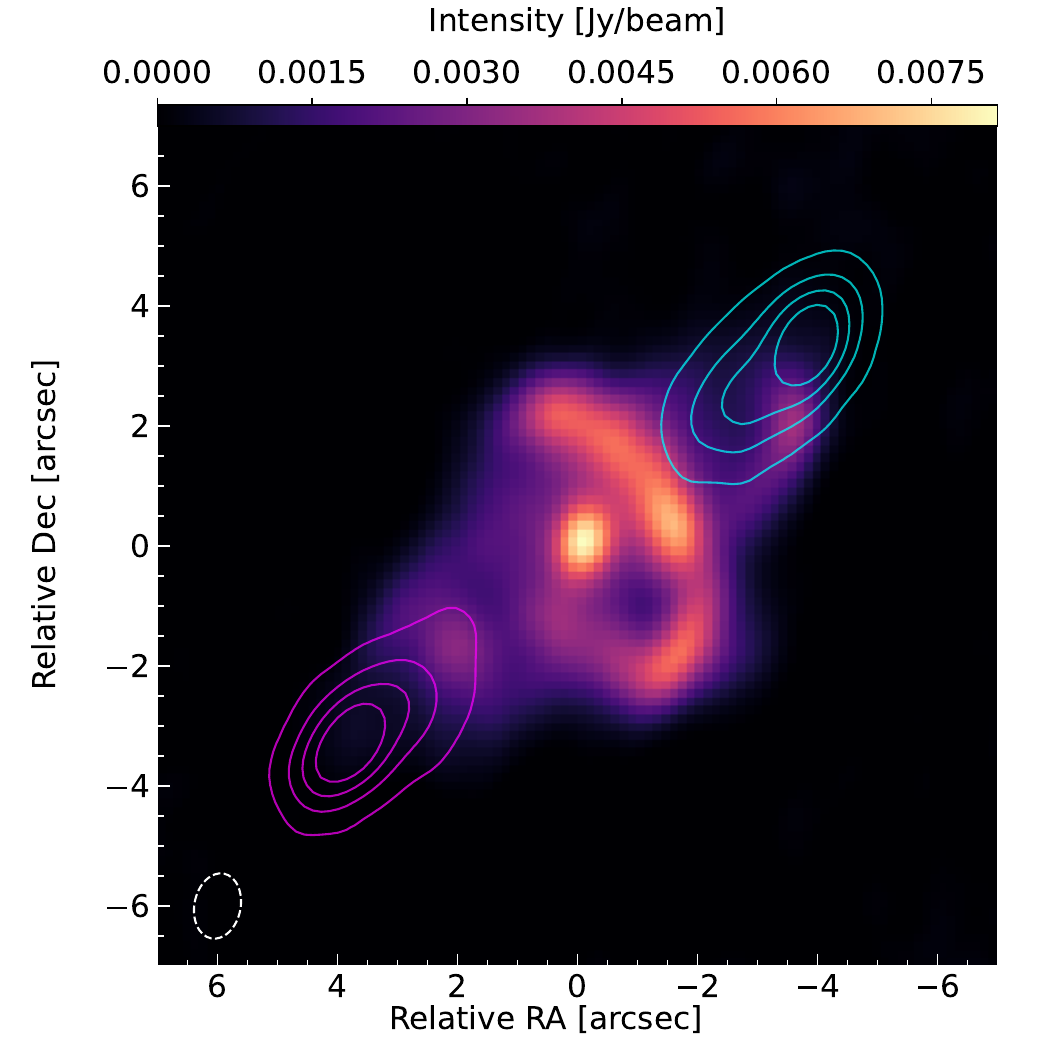}
\includegraphics[width=0.33\textwidth,clip, trim = 20 10 2 50]{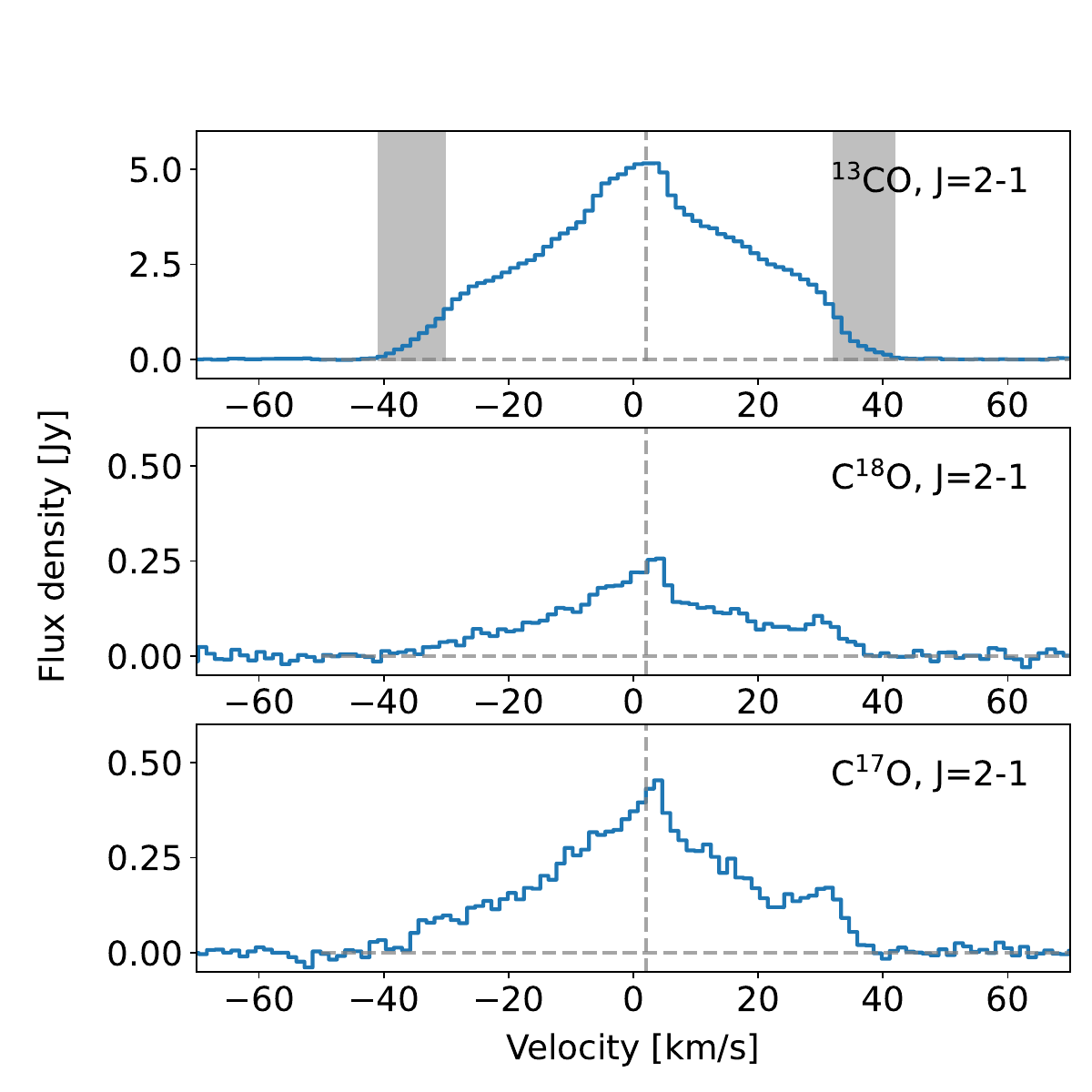}
 \caption{Continuum (1.3~mm) and \jtwoone\, $^{13}$CO, C$^{18}$O, and C$^{17}$O line emission towards M1-92.
 {\it Top:} Maps of the continuum (colour map) and high-velocity line emission (contours).
  The continuum peaks at $\alpha = \ra{19}{36}{18}{925}$ and $\delta = \dec{+29}{32}{49}{687}$.
The full lines show
 emission in the $^{13}$CO~\jtwoone\, integrated over the spectral intervals marked in grey in the line plot. Magenta and cyan contours show
 red-shifted and blue-shifted emission, respectively, at 70\%, 50\%, 30\%, and 10\% level of the
 peak values of 0.115 (red-shifted)
 and 0.127 (blue-shifted) ~${\rm Jy~beam^{-1}\times km~s^{-1}}$.
 {\it Bottom:} source-integrated spectra of the $^{13}$CO~\jtwoone, C$^{18}$O~\jtwoone, and C$^{17}$O~\jtwoone\, lines.}
   \label{fig:M1-92_highVel}
\end{center}
\end{figure}

The continuum image and the profiles of the $^{13}$CO, C$^{17}$O, and C$^{18}$O~\jtwoone\, lines extracted towards M1-92
using a region equal to the 3-$\sigma$ contour of the $^{13}$CO~\jtwoone\, line are shown in Fig.~\ref{fig:M1-92_highVel}.
The fluxes of the \jtwoone\, lines of $^{13}$CO, C$^{17}$O, and C$^{18}$O integrated over the same region are given in Table~\ref{tab:meas}.

\subsubsection{Continuum}

The continuum emission shows a complex morphology with many different components. In light of this,
we did not fit a Gaussian source to the continuum maps. We obtain an integrated flux density of $130 \pm 13$~mJy by integrating the emission over
the 3-$\sigma$ emission contour of the $^{13}$CO\,\jtwoone\, line. This is very similar to the value of $127 \pm 13$~mJy obtained from
integrating the emission over a circular region with diameter of 12\arcsec. The observed value
agrees within uncertainties with previous single-dish measurements
\citep[between 90 and 120 mJy][]{SanchezContreras1998,SanchezContreras2017}. 
Hence, no loss of flux is caused by the relatively large extent of the continuum emission.

\subsubsection{Line emission}

The line peaks indicate a systemic velocity of $\sim 2$~km~s$^{-1}$. The lines are about 85~km~s$^{-1}$ broad and tentatively display two main components,
a central one spanning a range of velocities $|\upsilon -  \upsilon_{\rm LRS}| \lesssim 10~{\rm km}$~s$^{-1}$ which appears on top of
the broader emission. The central component appears stronger relative to the broader component in the $^{13}$CO line
than in those of the rarer isotopologues. However, the distinction between the two components
is not very clear, and might not reflect two distinct physical velocity structures.

Integrating the high-velocity end of the line profiles reveals two emission regions which peak just outside of the region from where the continuum arises,
and in the same region as extensions are observed in the continuum emission.

\section{Analysis}

Based on the observed { angular sizes of the emission regions and the strengths of the \jtwoone\, lines of $^{13}$CO, C$^{17}$O, and C$^{18}$O listed in Table~\ref{tab:meas},
we derive linear sizes of the C$^{18}$O emission region based on the assumed distances ($\Delta D_{3\sigma}^{\rm C^{18}O}$),
$^{17}$O/$^{18}$O ratios, circumstellar masses ($M_{\rm gas}$),
recent mass-loss rates ($\dot{M}$), and $^{13}$CO abundances ($f^{\rm ^{13}CO}$).
Based on the $^{17}$O/$^{18}$O ratios we obtain, we infer initial masses using the MONASH ($M^{\rm KL16}_{\rm init}$) and FRUITY ($M^{\rm FRUITY}_{\rm init}$) models,
maximum luminosities using the evolutionary models of \cite{MillerBertolami2016} ($L_{\rm PAGB}$),
and maximum distances given these luminosities ($d_{\rm max}$). These derived quantities are given in Table~\ref{tab:isot}.}
For these calculations, we assume optically thin emission and
an excitation temperature, $T_{\rm exc}$, of 10~K for all three isotopologues of CO. 
The value of the excitation temperature is motivated by the low values usually observed towards
post-AGB stars with high molecular content \citep{Bujarrabal2001}.

For the calculation of the $^{17}$O/$^{18}$O ratios, total circumstellar masses and $^{13}$CO abundances, we used the line fluxes extracted from the regions enclosing the
3-$\sigma$ contour of the $^{13}$CO~\jtwoone\, moment-zero maps for each source as discussed above. The line fluxes obtained from these regions are
referred to as $S^{\rm ^{13}CO}$, $S^{\rm C^{17}O}$, and $S^{\rm C^{18}O}$.
For the calculation of the mass-loss rates at the end of the AGB ($\dot{M}$), we measure the C$^{18}$O~\jtwoone\, line fluxes extracted from smaller regions defined by the 3-$\sigma$ contours of the
C$^{18}$O~\jtwoone\, line (Fig.~\ref{fig:C18O}), with sizes $\Delta \theta_{3\sigma}^{\rm C^{18}O}$ (Table~\ref{tab:meas}).

 \begin{table*}
\centering
\caption{Observed fluxes of the $J=2-1$ lines and angular sizes of the emission region ($\Delta \theta_{3\sigma}^{\rm C^{18}O}$) and full widths at zero intensity ($\Delta\upsilon^{\rm C^{18}O}$)
of the C$^{18}$O~$J=2-1$ line.}
\label{tab:meas}
\begin{tabular}{ l c c c c c c c c }
Source &  $S^{\rm ^{13}CO}$ & $S^{\rm C^{17}O}$ & $S^{\rm C^{18}O}$ & $S_{3\sigma}^{\rm C^{18}O}$
& $\Delta \theta_{3\sigma}^{\rm C^{18}O}$ &  $\Delta\upsilon^{\rm C^{18}O}$\\
& [Jy~km~s$^{-1}$] & [Jy~km~s$^{-1}$] & [Jy~km~s$^{-1}$] & [Jy~km~s$^{-1}$] & [$\arcsec$] & [km~s$^{-1}$]\\
\hline
GLMP~950 & 15.3 (1.6) & 1.9 (0.3) & 1.13 (0.15) & 1.03 (0.14) & $2.4 \times 2.4$ & 41 \\
GLMP~953 & 23.3 (2.5) & 2.7 (0.3) & 2.1 (0.3) &  1.9 (0.3) & $3.0\times2.5$ & 37 \\
AFGL~5385 & 49 (5) & 2.3 (0.4) & 3.6 (0.5) & 3.8 (0.5) & $4.2\times3.5$ & 45 \\ 
HD~187885  & 64 (7) & 1.5 (0.3) & 2.4 (0.3) & 2.2 (0.3) & $3.2\times3.2$ & 61 \\ 
Hen~3-1475 & 39 (4) & 2.8 (0.3) & 2.2 (0.3) & 1.7 (0.3)  & $2.8\times2.4$ & 100 \\
M1-92 & 210 (20) & 16 (2) & 8.9 (1.3) & 7.6 (1.3) & $8.6 \times 4.4$ & 89 \\ 
\end{tabular}
\tablefoot{The uncertainties shown here include those from flux-calibration, which were not considered when calculating line-ratio uncertainties.}
\end{table*}

\subsection{$^{17}$O/$^{18}$O isotopic ratios}
\label{sec:isot}

We use the observed fluxes of the \jtwoone\,
lines of C$^{17}$O and C$^{18}$O to infer the $^{17}$O/$^{18}$O ratio. For this, we assume that the $^{17}$O/$^{18}$O
 isotopic ratio is equal to the C$^{17}$O/C$^{18}$O abundance ratio.
Given the low value assumed for the excitation temperature (10~K) and the slightly different energy levels of C$^{17}$O and C$^{18}$O, the fractions of
molecules in the $J=2$ level differ between the two isotopologues. Considering the excitation effects and the different Einstein-$A$ coefficients and
central frequencies of the lines ($\nu_\circ$), the $^{17}$O/$^{18}$O ratio can be
derived from the line flux ratio using
\begin{equation*}
\frac{^{17}\rm O}{^{18}\rm O}= \frac{S^{\rm C^{17}O}_{J=2-1}} {S^{\rm C^{18}O}_{J=2-1}} \frac{[A \nu_\circ]_{J=2-1}^{\rm C^{18}O}}{[A \nu_\circ]_{J=2-1}^{\rm C^{17}O}} 
\frac{Q^{\rm C^{17}O}}{Q^{\rm C^{18}O}} e^{(E^{\rm C^{17}O}_{J=2}-E^{\rm C^{18}O}_{J=2})/kT_{\rm exc}} \approx 0.95 \frac{S^{\rm C^{17}O}_{J=2-1}} {S^{\rm C^{18}O}_{J=2-1}}
\\
\end{equation*}
where $S$, $Q$, $E$, and $k$ are the frequency-integrated line flux density, the partition function, the energy
of the upper state, and the Boltzmann constant, respectively.
The small difference between the energies of the $J=2$ levels of C$^{17}$O and C$^{18}$O implies that assuming an excitation temperature
of 40~K instead of 10~K causes the derived abundance ratio to be only $\sim 3\%$ lower for a given observed line ratio. Hence, the isotopic
ratios we obtain are virtually independent of the assumed excitation temperature as long as the excitation of both molecules share
the same value.
The derived isotopic ratios are given in Table~\ref{tab:isot}.

\subsection{Circumstellar gas masses, recent mass-loss rates, and $^{13}$CO abundances}
\label{sec:mdot}

The molecular abundance with respect to hydrogen is a critical parameter required to derive the hydrogen gas masses in the CSEs from observed molecular lines.
Since the abundances of the CO isotopologues at the end of the AGB vary significantly as a function of stellar mass,
we selected the isotopologue with the smallest expected change for this calculation, C$^{18}$O.
For stars with initial masses between 1 and 4~$M_\odot$, the final $^{18}$O abundance is at
most $\sim 30\%$ lower than the initial value for the models of \cite{Karakas2016},
with more massive stars reaching larger differences than lower-mass ones.
For comparison, the final $^{13}$C and $^{17}$O abundance can vary by an order of magnitude or more from the initial value.
For the circumstellar mass calculations, we assume an average value of the
$^{18}$O (and C$^{18}$O) abundance, $f^{\rm C^{18}O}$, of $8.5 \times 10^{-7}$ relative to the number of hydrogen nuclei \citep{Karakas2016},
which is based on an initial solar abundance of $\sim 1.1\times 10^{-6}$. For HD~187885, which has been determined to
have a lower metallicity ($Z=0.0035$), we adopt an abundance of $2.1 \times 10^{-7}$, also based on the models presented by \cite{Karakas2016}.
The total circumstellar gas mass is calculated using the following expression
\begin{equation}
\label{eq:Mgas}
M_{\rm gas} = \frac{M_{\rm H}}{0.7} = \frac{1}{0.7}\frac{4~\pi~d^2~Q^{\rm C^{18}O}~m_{\rm H}}{g_{J=2}~A~h~\nu_\circ~f^{\rm C^{18}O}} e^{\frac{E_{J=2}}{T_{\rm exc}}}~S^{\rm C^{18}O},
\end{equation}
where $d$ is the distance to the source, $g_{J=2}$ is the statistical weight of the $J=2$ level, and $M_{\rm H}$ and $m_{\rm H}$ are the total hydrogen mass and the mass of a hydrogen atom, respectively. 
The factor 0.7 corresponds to the approximate fraction of the total mass accounted for by hydrogen at the end of the AGB \citep{Karakas2016}.
The derived circumstellar gas masses are given in Table~\ref{tab:isot}.

The recent mass-loss rates are determined using Eq.~\ref{eq:Mgas} with $S^{\rm C^{18}O}$ replaced by $S^{\rm C^{18}O}_{3\sigma}$.
This approach is chosen because most of the C$^{18}$O~\jtwoone\, emission is produced in smaller regions (Fig.~\ref{fig:C18O}),
than those defined by the 3-$\sigma$
$^{13}$CO~\jtwoone\, contours (Fig.~\ref{fig:13CO}).
The kinematic ages ($\Delta t$) of the C$^{18}$O~\jtwoone\, emission regions are also determined using the 3-$\sigma$ C$^{18}$O~\jtwoone\, contours.
The line widths used for the calculation of the kinematic ages of the C$^{18}$O~\jtwoone\, emission regions
correspond to the widths of the C$^{18}$O~\jtwoone\, line profiles at zero intensity.
Using the derived expansion velocity, the size of the emission region and
the adopted distances, we determined kinematic ages
of the C$^{18}$O~$J=2-1$ emission  ($\Delta t$). Based on these ages and on the gas masses within the 3-$\sigma$ C$^{18}$O~\jtwoone\, regions,
we determine a recent mass-loss rate for each source.
These mass-loss rates represent lower limits to the maximum mass-loss rates experienced by each source recently,
because the phase of high AGB mass loss probably ceased even hundreds of years ago for some sources (see Section~\ref{sec:disc_mdot}).

Using an analogous expression to Eq.~\ref{eq:Mgas},
we also derive the $^{13}$CO abundance relative to hydrogen nuclei ($f^{\rm ^{13}CO}$) that would make the estimated gas masses consistent with those derived based on
the C$^{18}$O~\jtwoone\, line. Although we assume optically thin emission,
optical depth effects seem to affect the $^{13}$CO emission to some extent because for all sources the brightness temperatures
observed towards the peak of the integrated emission reach values (between 7 and 13~K)
comparable to the excitation temperature we assume.
If our assumed excitation temperature
is correct, this implies optical depths $\gtrsim 1$.
Nonetheless, the relatively high values of the
$^{13}$CO abundances we derive (see below) suggest that the optical depth effects are not very strong.

\section{Discussion}
\label{sec:disc}

\subsection{Isotopic ratios}

The $^{17}$O/$^{18}$O isotopic ratios allow us to estimate the initial masses of the sample stars using results from stellar evolution models.
These are powerful constraints for understanding the properties and evolution of these sources, and allow us to
infer stellar luminosities and distances and investigate whether the sources with known spectral types have evolved within the expected timescales.

\subsubsection{Comparison to other sources}

To put the values of the $^{17}$O/$^{18}$O ratios determined for the sample stars in context, we
compare them to those available in the literature for AGB stars using the sample of C-rich AGB stars studied by \cite{Abia2017},
with a total of 29 objects, and 
the samples of O-rich AGB stars studied by \cite{Hinkle2016} and
\cite{Lebzelter2019}, which include a total of 80 sources.

The minimum and maximum values reported by \cite{Abia2017} for C-rich sources are $\sim 0.5$ and $\sim 2.5$. Only three sources ($\sim 10\%$) display
ratios $\lesssim 0.6$. Hence, the values we obtain for AFGL~5385 (0.61) and HD~187885 (0.59) do occur
in AGB stars as well, but are relatively rare. The values we report for
GLMP~950 and GLMP~953 are very typical for C-rich AGB stars based on the sample of \cite{Abia2017}. The value of the $^{17}$O/$^{18}$O ratio is not expected
to change during the AGB or post-AGB for stars that become rich
in carbon and do not experience HBB. Hence, comparing the values for post-AGB and AGB C-rich stars is
meaningful. Nonetheless, we note that a direct comparison of the observed distributions is not straight-forward, because the
duration of the carbon-rich phase in the AGB and that of the post-AGB phase depend strongly on initial mass and metallicity.

The $^{17}$O/$^{18}$O ratios reported by \cite{Hinkle2016} and \cite{Lebzelter2019} for O-rich AGB stars cover a broad range
of values, from $\sim 0.05$ to $\sim 16$. Nonetheless, extreme values are relatively rare, with only $\sim 11\%$ having ratios lower than 0.2
and $\sim 11\%$ having values larger than 2. Hence, the values we obtain for Hen~3-1475 (1.29) and M1-92 (1.8) are typical for O-rich
AGB stars. However, we emphasize that  these ratios were determined at different points during the AGB for each source,
and at least some of the O-rich stars studied by \cite{Hinkle2016} and \cite{Lebzelter2019} will become C-rich or experience HBB
before entering the post-AGB phase.

\begin{table*}
\centering
\caption{ Quantities derived from the observed values given in Table~\ref{tab:meas}.}
\label{tab:isot}
\begin{tabular}{ l c c c c c c c c c c}
Source & $\Delta D_{3\sigma \tablefootmark{\dag}}^{\rm C{18}O}$ & $M_{\rm gas}\tablefootmark{\ddag}$ & $f^{^{13}\rm CO}$ & $\Delta t$ & $\dot{M}$ & $^{17}$O/$^{18}$O & $M^{\rm KL16}_{\rm init}$ & $M^{\rm FRUITY}_{\rm init}$
& $L_{\rm PAGB}$ & $d_{\rm max}$\\
&  [au] & [$M_\odot$] & [$\times10^{-5}$] & [yr] & [$M_\odot~{\rm yr}^{-1}$] & & [$M_\odot$] & [$M_\odot$] & [$10^3~L_\odot$] & [kpc] \\
\hline
GLMP~950  & 9360 & 0.10  & 1.1\phantom{$^\ddagger$}  & 1085 & $\phantom{1.}8\times10^{-5}$ & 1.7 (0.2) & 1.8 & $\phantom{<~} 1.8$ & $\phantom{\sim~} 6.9$ & $\phantom{\sim~} 4.2$ \\ 
GLMP~953 & 11400 & 0.17 & 0.9\phantom{$^\ddagger$} & 1460 & $1.1\times10^{-4}$ & 1.3 (0.2)  & 1.7 & $\phantom{<~} 1.7$ & $\phantom{\sim~} 6.7$ & $\phantom{\sim~} 4.0$ \\ 
AFGL~5385 & 14280 & 0.24 & 1.1\tablefootmark{$\ast$} & 1510 & $1.7\times10^{-4}$ & 0.61 (0.10) & 1.4 & $\phantom{<~} 1.5$ & $\phantom{\sim~} 5.9$ & $\phantom{\sim~} 2.4$ \\ 
HD~187885$^ \dag$ & 8000 & 0.35 & $\phantom{^\star}$0.55\tablefootmark{$\ast$} & 620 & $\phantom{1.}5\times10^{-4}$ & 0.59 (0.12)  & $1.15$ & $< 1.3$ & $\sim 6.2$ & $\sim 2.5$ \\ 
Hen~3-1475 & 16240 & 0.42  & 1.5\phantom{$^\ddagger$}  & 770 & $\phantom{1.}4\times10^{-4}$ & 1.29 (0.15) & 1.7 & $\phantom{<~} 1.7$ & $\phantom{\sim~} 6.7$ & $\phantom{\sim~} 4.5$ \\ 
M1-92  & 22360 & 0.35 & 2.0\phantom{$^\ddagger$}  & 1190 & $2.5\times10^{-4}$ & 1.8 (0.2) & 1.8 & $\phantom{<~} 1.8$ & $\phantom{\sim~} 6.9$ & $\phantom{\sim~} 2.8$ \\ 
\end{tabular}
\tablefoot{\tablefoottext{\dag}{Using the larger value of $\Delta \theta_{3\sigma}^{\rm C^{18}O}$.}
      \tablefoottext{\ddag}{For an assumed excitation temperature of 10~K and C$^{18}$O abundance relative to the number of hydrogen nuclei of $2.1 \times 10^{-7}$ for HD~187885 and $8.5 \times 10^{-7}$ for the other sources (see Section~\ref{sec:mdot}).} \tablefoottext{$\ast$}{These values might be underestimated because
      the corresponding $^{13}$CO~\jtwoone\, line seems to be affected by optical depth effects.}}

\end{table*}

Observations of isotopic ratios in post-AGB stars are much rarer, and to our knowledge
only six C-rich and ten O-rich sources have had $^{17}$O/$^{18}$O
ratios determined thus far \citep[Alcolea et al. in prep.]{Khouri2022,GallardoCava2022}. These samples are also heavily biased by observations of specific subgroups of post-AGB stars,
such as the eight O-rich water fountain sources studied by \cite{Khouri2022}. The observed values range roughly from 0.3 to 4 for C-rich objects
and from 0.2 to 6 for O-rich ones. Hence, the values reported by us are within the observed ranges.

\subsubsection{Derived initial masses}

We compare the isotopic ratios derived in this work to stellar evolution models using the MONASH code
\citep{Karakas2016,Karakas2018} and the FRUITY database \citep{Cristallo2015}.
These models assume a solar-like composition scaled to different metallicities.
While there is no strong disagreement between the predictions of the models regarding the first dredge up, which has the dominant effect
on the oxygen isotopic ratios for stars with initial masses $\lesssim 4$~$M_\odot$,
the effects of the third dredge up are much more uncertain \citep[e.g., ][]{Palmerini2021}.
The uncertain aspects relative to the third dredge-up
relevant for the analysis presented here concern the mass ranges for which stars become carbon rich
and for which HBB destroys $^{18}$O in the stellar envelope efficiently.

The observed isotopic ratios imply remarkably similar initial masses for a given metallicity for the two models. The derived values are given in
Table~\ref{tab:isot} for solar metallicity, with the exception of HD~187885 for which we consider a metallicity $Z=0.0035$ \citep{vanWinckel2000}.
M1-92 is the only object with a previous determination of the $^{17}$O/$^{18}$O ratio. Our value of $1.8\pm0.1$ is in agreement within
uncertainties with the value reported in the literature \citep[$1.6\pm0.15$, ][]{Alcolea2022}. The value that we obtain
implies an initial mass $\sim 1.8~M_\odot$ for solar composition, similar to the value obtained by \citeauthor{Alcolea2022}, $\sim 1.7~M_\odot$.

While the MONASH and FRUITY models with the exact metallicity of HD~187885 ([Fe/H] = -0.6, or $Z=0.0035$)
are not available in the literature, we can conclude that
the predicted trends in the $^{17}$O/$^{18}$O ratio with metallicity imply an initial mass smaller than $<1.3~M_\odot$ in both cases.
In the case of the MONASH code, the predicted $^{17}$O/$^{18}$O ratios imply an initial mass of $\sim 1.15~M_\odot$ for $Z=0.0028$ \citep{Karakas2018},
which is the closest metallicity to that of HD~187885 for which the models have been calculated. For the FRUITY database, the models
with lowest initial mass available ($1.3~M_\odot$) predict larger $^{17}$O/$^{18}$O values for both $Z=0.003$ and $Z=0.006$,
and the predicted value decreases with decreasing initial mass in this range. The relatively low mass and carbon-rich nature of this source is
in agreement with the findings from \cite{Karakas2018}, which imply that stars with masses as low as 1.15~$M_\odot$ become
rich in carbon by the end of the AGB for $Z=0.0028$. Hence, the initial mass of HD~187885 seems to be very close to the minimum
mass required for becoming a carbon star given its initial metallicity. We note that the models of \cite{Karakas2018} assume initial isotopic ratios
equal to solar even at lower metallicities, which might not be representative of the initial isotopic ratios of HD~187885. 
Our results imply initial masses for GLMP~950 and GLMP~953 (1.8 and 1.7~$M_\odot$, respectively)
within the range usually attributed to carbon stars.

Interestingly, the derived initial masses for Hen~3-1475 and M1-92 also fall in the range expected for stars to become carbon rich. The fact that
these sources are observed to still be oxygen rich in the post-AGB phase might be explained by an interrupted
evolution, { for instance with the envelope being ejected by interactions with a close companion} \citep[e.g., ][]{Khouri2022,Alcolea2022}.
Alternatively, these sources might have an initial metallicity higher than solar and
initial masses between  2.0 to 2.4~$M_\odot$. In this scenario, their O-rich nature could be consistent with stellar evolution models,
which remain oxygen rich up to initial masses $\sim 2.4~M_\odot$ at higher metallicity (e.g. $Z=0.03$).

{ Despite the encouraging agreement between the two evolutionary models considered for the interpretation of the derived $^{17}$O/$^{18}$O ratios, we note that the initial
masses we derive are fairly uncertain both because of uncertainties in the model calculations and in the initial composition of these sources.
More robust results can be obtained using larger statistical samples.}

\subsubsection{Implications for luminosities and distances}

The initial masses we derive are relatively low, and set limits on the maximum luminosities reached by each star. To assess how these
limits compare to our assumptions, we determine the maximum expected post-AGB luminosity for each source from
evolutionary models presented by \cite{MillerBertolami2016} given the inferred initial masses.
Using these luminosities (presented in Table~\ref{tab:isot}),
we scaled the distance for each sources based on the luminosities adopted from the literature (Table~\ref{tab:cont}).
These distances are also given in Table~\ref{tab:isot}.
The distances for GLMP~950, GLMP~953, HD~187885, and M1-92 are consistent with
those adopted from the literature with differences $< 10\%$, while those for AFGL~5385 and Hen~3-1475 show larger discrepancies
of 30\% and 22\%, respectively. Nonetheless, considering the uncertainties in the derived initial masses and on the determination of distances, the overall agreement is
good. We note that distances up to 8.3~kpc have been reported for Hen~3-1475 \citep{Borkowski2001}.
Our results indicate that such large distances are not realistic. { \cite{Riera2003} derived a distance of 5.8~kpc based
on improved measurements of radial velocities and proper motions of the jet features. Our results suggest an even smaller distance which, based on the discussion in \cite{Riera2003},
implies a larger value $(\gtrsim 55^\circ)$ for the inclination angle of the jet than what is assumed in their calculations ($50^\circ$).}

\subsubsection{Implications for post-AGB lifetimes}

{ Given the definition of the beginning of the post-AGB as} when the mass of the envelope becomes 0.01 times the present-day mass of the given star
\citep[e.g.][]{MillerBertolami2016}, the large
circumstellar masses we derive imply that these sources { must have been} on the AGB when the high-mass-loss phase { started}.
Hence, the kinematic ages given in Table~\ref{tab:isot} represent an upper limit for the duration of the post-AGB evolution of these sources up to now.
Using the spectral types available in the literature for four of the sources
(AFGL~5385, HD~187885, Hen~3-1475, and M1-92, see Table~\ref{tab:cont}) we are able to compare the time
these objects took to transition between the end of the AGB and their present position in the Hertzsprung–Russell (HR) diagram
with predictions from evolutionary models.

Relatively large uncertainties are associated with
the timescales predicted by stellar evolution models mostly because of the unknown evolution of the mass-loss rate in the post-AGB phase.
This is particularly critical in the beginning of the post-AGB, because stellar temperatures are still relatively low in this phase and AGB-like mass loss can still take place
\citep{MillerBertolami2016}.
\citeauthor{MillerBertolami2016} divides the evolution during the post-AGB phase in two stages, the early and late post-AGB, with associated
timescales $\tau_{\rm tr}$ and $\tau_{\rm cross}$, respectively. The final point of the post-AGB evolution is defined when the maximum effective temperature
is reached, which corresponds to $T_{\rm eff} > 10^5$~K in all models presented by \cite{MillerBertolami2016}.

Since the changes experienced by the stars are gradual, definitions
of transition points are to a large extent arbitrary. Nonetheless, the general trend is that model stars move to the blue
in the HR diagram as the envelope mass decreases. This change is slow at first and accelerates strongly as the envelope mass becomes lower.
At the start of the post-AGB (envelope mass equal to 0.01 of present-day mass),
the effective temperature of the model stars are relatively low ($\lesssim 4000$~K) for initial stellar mass between 1.2 and 2~$M_\odot$,
as inferred from Fig.~8 in \cite{MillerBertolami2016}.
The point of transition between early and late post-AGB phase is set by \cite{MillerBertolami2016}
to be log($T_{\rm eff}$) = 3.85 ($T_{\rm eff} \sim 7080$~K). 
Since the change in temperature as a function of time is fast in the late post-AGB, the definition of this transition
point is of relative little consequence for the derived timescales according to the author,
as long as the temperatures are relatively low but high enough to prevent AGB-like outflows.

AFGL~5385, HD~187885, Hen~3-1475, and M1-92 have
effective temperatures ranging between $\sim 6800$~K and $\sim 30000$~K based on their spectral types (Table~\ref{tab:cont}).
Hence, they have evolved significantly beyond what would be considered
the start of the post-AGB phase. Moreover, only the source with the lowest effective temperature (AFGL~5385)
would be classified as an early post-AGB source,
but even this object is relatively close to transitioning into the late post-AGB in which a rapid evolution of the effective temperature takes place.

By comparing the predicted times for stars to evolve from the beginning of the post-AGB
to effective temperatures $\sim 7000$~K ($\tau_{\rm tr}$) to the kinematic ages of the C$^{18}$O~$J=2-1$ emission,
it is clear that these four sources had an evolution faster than predicted by models by a factor of a few, at least.
By interpolating the values for $Z=0.02$ and $Z=0.01$
in Table~3 in \cite{MillerBertolami2016}, we find that $\tau_{\rm tr}$ varies between $\sim 4400$~yr for 1.25~$M_\odot$ stars
to $\sim 2370$~yr for 2.0~$M_\odot$ for solar metallicity ($Z=0.014$). At lower metallicities ($Z=0.001$), $\tau_{\rm tr}$ becomes shorter and
varies between $\sim 3700$~yr and $\sim 1600$~yr for models with 1.15 and 2.0~$M_\odot$.
AFGL~5385, HD~187885, Hen~3-1475, and M1-92 all seem to have reached their present position at least two times faster (and even faster in most cases)
than expected. This particularly striking for M1-92 and Hen~3-1475 which have B-type central stars, with $T_{\rm eff} > 10000$~K, and kinematic ages $\lesssim 1000$~yr.
HD~187885 also seems to have evolved more than five times faster than expected into the late post-AGB phase.

{ Previous models for post-AGB evolution presented by \cite{Vassiliadis1994} display on average a slower evolution
between the end of the strong AGB-like mass loss
and the less intense post-AGB one. However, the comparison between different stellar evolution models is not completely straight forward because of different definitions of transitions points.
In contrast to \citeauthor{MillerBertolami2016}, \cite{Vassiliadis1994} define the beginning of the post-AGB phase
as when the stellar temperature in the models increases by a factor of two from the value during the AGB.
For solar composition,
this results in stellar temperatures at the beginning of the post-AGB varying between $\sim 4500$~K and $\sim 4200$~K for star with initial masses between
1.5 and 2~M$_\odot$, respectively. From these starting points, the models reached stellar temperatures of $10^4$~K in $\gtrsim 8000$~yr for initial masses
$\gtrsim 1.5~{\rm M}_\odot$. For a metallicity $Z=0.004$ comparable to that of HD~187885 ($Z=0.0035$)
and an initial mass equal to 1~$M_\odot$, the evolution is shorter $\sim 6000$~yr from an initial post-AGB stellar temperature $\sim 5600$~K.}

{  The timescales reported by \cite{Vassiliadis1994} are
almost an order of magnitude larger than that inferred for the evolution of Hen~3-1475 and M1-92, which have evolved to stellar temperature larger than 10$^4$~K.
A comparison to HD~187885 and AFGL~5385 is not as clear because
these sources have stellar temperatures lower than 10$^4$~K, and the durations of shorter steps of evolution below this limit are not specified.
According to the models,
 AFGL~5385 and HD~187885 should have had stellar temperatures at the beginning of the post-AGB phase of $\sim 4500$~K and $\sim 5600$~K, respectively.
The evolution of AFGL~5385 to the present value of $\sim 7300$~K in 1500~yr seems to have happened faster than predicted
considering that the source has experienced half the evolution to 10$^4$~K (expected to happen in $\sim 8600$~yr)
and that the pace of evolution is slower at lower stellar temperatures. For HD~187885, the change in temperature to the present value of $\sim 6800$~K
is relatively small, and we refrain from making a comparison that would require too many assumptions.}

The fast evolution { we infer with respect to those predicted by models discussed above} might be caused by a different evolutionary path from the
single-star evolution considered in the models. However, if at least some of these four
sources evolved without the direct effect of a close companion, our results indicate that the central remnant warms up faster than expected.
\cite{SanchezContreras2017} also suggested a faster evolution of the central source in the post-AGB phase
based on high mass-loss rates derived for a small sample of post-AGB sources.

\subsection{$^{13}$CO and C$^{18}$O abundances}

For the stars assumed to have initial solar composition,
we find $^{13}$CO abundances between $0.9\times 10^{-5}$ and \mbox{$2.0\times 10^{-5}$} with respect to the total number density of hydrogen
atoms (including atomic and molecular sources). We find the largest values of 1.5 and $2.0\times10^{-5}$ for Hen~3-1475 and M1-92,
which are oxygen-rich sources. This fits the expectation that the abundance of $^{13}$C is larger in O-rich stars, but the values are 1.5 and
2 times larger than often assumed for post-AGB stars with solar-like elemental abundances and high molecular content, $\sim 1\times10^{-5}$ \citep[e.g., ][]{Bujarrabal2001,Bujarrabal2017}.
The values for C-rich sources with assumed solar metallicity are lower with respect to the O-rich ones, but the values for these five sources
differ only by roughly a factor of two.
The relatively high abundances of $^{13}$C we find are somewhat surprising
because optical depth effects are expected to affect the $^{13}$CO lines more strongly than the C$^{18}$O ones, which should lead to lower
inferred values for the $^{13}$CO abundances than the real ones. From this, it seems that optical depth effects are at most mild in the $^{13}$CO lines.

If the C$^{18}$O envelope is much smaller than the $^{13}$CO because of more efficient dissociation,
we would overestimate the $^{13}$CO abundances.
To examine this possibility, we compared the ratio between intensities of the $^{13}$CO
and C$^{18}$O~$J=2-1$
lines in the central beam of the maps to the same ratio between the line fluxes given in Table~\ref{tab:meas}, which are
integrated over larger apertures. We find larger ratios for all sources when larger apertures are used,
implying that the C$^{18}$O emission is more centrally peaked than the that of $^{13}$CO. This could be caused by
the $^{13}$CO emission being either optically thick in the central beam or more extended than that of C$^{18}$O.  
Nonetheless, compared to the ratios from $^{13}$CO~\jtwoone~ 3-$\sigma$ regions, those measured from the central beams are only between 15 and 20\% smaller
for GLMP~950, GLMP~953, HD~187885, and M1-92, and 30\% smaller for AFGL~5385 and
Hen~3-1475. Furthermore, the different sizes
of the  3-$\sigma$ contours of the \jtwoone emission of $^{13}$CO and C$^{18}$O are consistent with the lower signal-to-noise
in the maps of the C$^{18}$O lines. Based on these considerations,
we conclude that, if present, the effect of different dissociation radii on the line ratios seems to be relatively small, and the $^{13}$CO abundance
seems to be indeed relatively high in these sources.

Interestingly, the three C-rich sources with assumed solar metallicity display $^{13}$CO/C$^{18}$O ratios between 10 and 13, while the C-rich source with lower metallicity (HD~187885) shows
a value of $\sim 26$, which is even larger than the values of 18 (Hen~3-1475) and 24 (M1-92) determined for the two O-rich sources. This might be caused by the
chemical composition of HD~187885 not having been as strongly affected by the third dredge-up as those of the other
C-rich sources, given the relatively low value of its C/O ratio $\sim 1$ \citep{vanWinckel1996,vanWinckel2000}.

\subsection{Circumstellar gas masses}

\subsubsection{Uncertainties}
The uncertainties on the circumstellar masses determined by us arise from uncertainties on the excitation temperature, the distance, and the C$^{18}$O
abundance. The circumstellar masses
depend on the assumed excitation temperature in a non-linear manner.
Increasing the excitation temperature from 10~K to 20~K causes the derived masses to decrease by approximately 20\%,
while increasing it from 10~K to 40~K causes the derived masses to increase by about 10\%. Hence, we estimate the uncertainty on the
excitation temperature translates to an uncertainty of at most $\sim 20\%$ on the derived circumstellar gas masses.
The circumstellar masses are proportional to the square of the adopted distance,
which we assume to be uncertain by 30\%. Considering the uncertainty of 30\% from $^{18}$O abundance variation
(Section~\ref{sec:mdot}), we find the circumstellar masses to be uncertain by a
factor of 1.7. This does not account for uncertainties regarding the initial metallicity or, more specifically, the initial $^{18}$O abundance.

\subsubsection{Comparison to the literature}

The derived circumstellar gas masses correspond to a substantial fraction of the initial stellar mass
of each source, given the initial masses $< 2~M_\odot$. 
These values are on the high end when compared to those observed in a sample of 21~ post-AGB stars, pre-PN and PN,
for which circumstellar masses between $\sim 10^{-2}$ and 1~$M_\odot$ were inferred \citep{SanchezContreras2012}.

\cite{Oppenheimer2005} modelled the spectral energy distribution of AFGL~5385 and derived a dust mass-loss rate of $\sim 10^{-6}~M_\odot~{\rm yr}^{-1}$.
This is somewhat high compared to gas mass-loss rate of $1.7 \times 10^{-4}~M_\odot~{\rm yr}^{-1}$ derived by us, and would imply a gas-to-dust ratio of
170. This relatively low value implies substantial depletion of refractory elements, but is comparable
to values $\sim 200$ determined for AGB stars with large mass-loss rates $\gtrsim 10^{-5}~M_\odot~{\rm yr}^{-1}$
\citep[e.g.][]{Blommaert2018,Olofsson2022}. The outer radius derived from their model implied a timescale for building up the envelope larger than 2000~yr
for an assumed expansion velocity of 15~km~s$^{-1}$. The expansion velocity we find using the C$^{18}$O~\jtwoone\, line (22.5~km~s$^{-1}$) would imply
a timescale for the dust envelope larger than 1300~yr, which is consistent with the kinematic ages of the C$^{18}$O~\jtwoone\, emission,
$\sim 1500$~yr.

The total circumstellar masses derived by \cite{Bujarrabal2001} of $\sim 2.6 \times 10^{-2}~M_\odot$ and 0.63~$M_\odot$ for HD~187885 and
Hen~3-1475, respectively, can be compared to the values we find by using the distances assumed by them
(1~kpc for HD~187885 and 5~kpc for Hen~3-1475). We find a circumstellar mass of $6 \times 10^{-2}$ and 0.32~$M_\odot$, for HD~187885 and Hen~3-1475, respectively.
The discrepancy is mostly explained by the different value of the
$^{13}$CO abundance relative to hydrogen adopted by  \cite{Bujarrabal2001}, $1\times10^{-5}$, compared to that derived by us,
$1.5\times10^{-5}$ and $0.55\times10^{-5}$ for
Hen~3-1475 and HD~187885, respectively.

A significantly larger circumstellar gas mass ($\sim 1~M_\odot$) has been determined for M1-92 \citep{Alcolea2022} than what we find ($0.35~M_\odot$). This can also be mostly explained by
a difference between the $^{13}$CO abundance assumed by \citeauthor{Alcolea2022} ($1\times10^{-5}$) and that determined by us ($2\times10^{-5}$).
\cite{Li2024} also find a circumstellar mass $\sim 1~M_\odot$ based on observations of optical scattered light and an assumed gas-to-dust ratio of 200. The lower circumstellar gas mass
we find would imply a lower gas-to-dust ratio $\sim 70$, which is quite low even if full condensation of refractory elements is considered.

\subsection{Recent mass-loss rates}
\label{sec:disc_mdot}

{ The mass-loss rates we derive $\sim 10^{-4}$~M$_\odot$~yr$^{-1}$ are roughly an order of magnitude higher than the peak values achieved by stars with initial
masses between 1 and 2~M$_\odot$ ($\sim 10^{-5}$~M$_\odot$~yr$^{-1}$) in commonly used mass-loss-rate prescriptions \citep[e.g.][]{Vassiliadis1993}. Nonetheless, our calculations
might even underestimate the actual peak mass-loss rates experienced by these sources. For instance,}
the inner radius of the envelope of HD~187885 observed by \cite{Lagadec2011}, of 0.4\arcsec, implies a timescale for the cessation of mass loss
$\sim 160$~yr ago assuming the same expansion velocity we use, $\sim 30$~km~s$^{-1}$.
This is only four times smaller than kinematic age of the C$^{18}$O~\jtwoone\, derived by us.
The dust model calculated by \cite{Oppenheimer2005} for AFGL~5385 also implies a cessation of the high-mass-loss-rate phase $\sim 300$~years ago (at the
distance of 3.4~kpc considered by us). Hence, { these non-negligible times since copious mass loss has ceased
imply lower times spent in the phase of large mass-loss rates than given in Table~\ref{tab:isot}, indicating that the maximum instantaneous
mass-loss rates experienced by these sources are even larger than the average values we report}.

We also note that the derived timescales pertain to the region traced by the \jtwoone\, line of the observed isotopologues. If the size of the emission region
is being determined by dissociation or excitation of the $J=2$ levels, rather than by the onset of the high-mass-loss-rate phase,
the timescales and circumstellar masses we infer will be lower than those experienced by the sources.
Nonetheless, the obtained mass-loss rates are independent of whether the high-mass-loss-rate phase
has been ongoing for longer than listed in Table~\ref{tab:isot}. 

\subsection{Progenitor sources}
\label{sec:prog}

One interesting question is whether the oxygen-rich stars in our sample could be an evolved stage of water fountain sources.
The inferred kinematic ages for the C$^{18}$O~$J=2-1$ emission for M1-92 and Hen~3-1475 ($\sim 1190$ and $\sim 770$~years, respectively) are a factor
of at least four larger than typical ages for the envelopes of water fountains \citep[$\lesssim 200$~years,][]{Khouri2022}.
When compared to average mass-loss rates derived for water fountains \citep[$\gtrsim 10^{-3}~M_\odot$~yr$^{-1}$, ][]{Khouri2022},
the values we present here are lower. However, this mostly reflects the larger timescales since the creation of the envelopes.
If the intense mass-loss process of M1-92 and Hen~3-1475 lasted $\lesssim 200$~years instead of the kinetic ages of the C$^{18}$O~\jtwoone\, we determine,
their mass-loss rates would have been $\sim 10^{-3}~M_\odot$~yr$^{-1}$, in the same range derived for water fountains.
In fact, \cite{Alcolea2022} argues that it is not possible to
exclude scenarios in which the whole envelope of M1-92 was created in a short mass-loss event. Moreover, the large excess momentum
in the outflow compared to that provided by the radiation field in these two sources \citep{Bujarrabal2001} imply an acceleration mechanism
that probably requires the action of a companion. \cite{Khouri2022} suggested
water fountains to have experienced common-envelope evolution, and similar suggestions have
been made for M1-92 \citep[e.g., ][]{Alcolea2022,Li2024} and
Hen~3-1475 \citep{Borkowski2001}.
The redder colours of water fountains (see~Fig.~\ref{fig:colour-colour}) are most likely caused at least in part by their envelopes being more compact,
with shorter timescales since the start of their high-mass-loss phase.
Moreover, the high-velocity water masers, which are the distinguishing feature of water fountains, are only expected to be produced in the $\sim 200$~years 
that define the class. Whether M1-92 and Hen~3-1475 displayed such high-velocity masers several hundred years ago is possible but unknown.

{ Confirming that these classes of sources are associated is not straight-forward, but statistical studies that include more sources at different evolutionary
stages and should reveal whether this scenario is in agreement with observations when taking into account the different relevant timescales. Moreover, hydrodynamical
modeling of the evolution of water fountain systems should also indicate whether the evolution of spectral energy distributions and morphologies are consistent.
Another interesting question that deserves further investigation is
whether an evolutionary relationship exists between post-AGB stars with dusty discs \citep[e.g.]{vanWinckel2007} and the objects studies here and water fountains.}

Regarding the carbon-rich stars, their expected too-low water abundances imply that water masers are not produced,
which would prevent these sources from even being classified as water fountains.
Nonetheless, their compact envelopes and high mass-loss rates suggests that sources like HD~187885, GLMP~950, GLMP~953, and AFGL~5385
might be carbon-rich stars that have taken the same evolutionary path
as that associated with water-fountain sources, i.e., common-envelope evolution.

\subsection{High-velocity outflow of Hen~3-1475}

\begin{figure}[h!]
\begin{center}
 \includegraphics[width=0.45\textwidth,clip,trim=13 90 28 100 ]{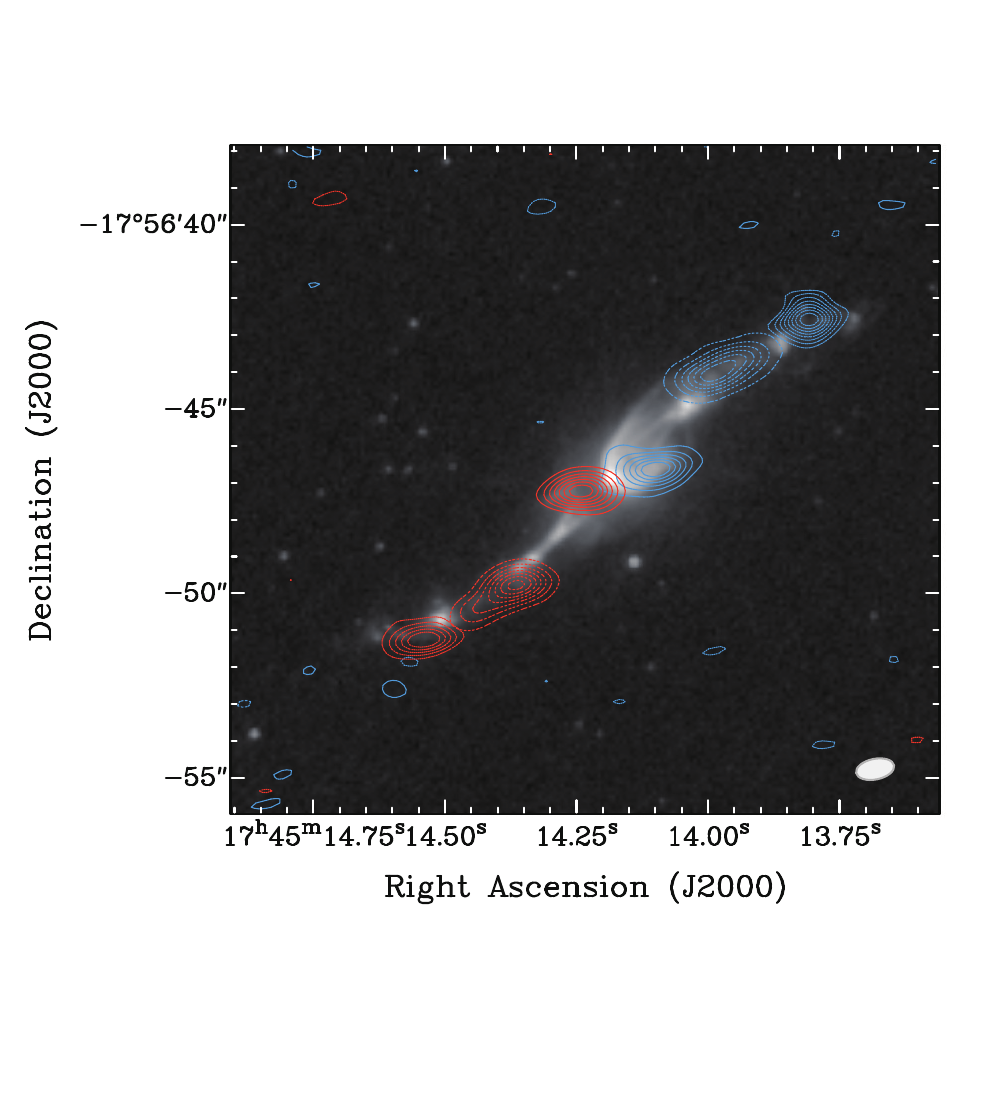} 
 \caption{Maps of the high-velocity $^{13}$CO~\jtwoone\, emission towards Hen~3-1475
 as shown in Fig.~\ref{fig:Hen_highVel} compared to images obtained with the {\it HST} WFPC2 through the F658N ([N II]
at 6584~$\AA$) filter \citep{Riera2003}.}
   \label{fig:Hen3-1475_Hubble}
\end{center}
\end{figure}

As shown in Fig.~\ref{fig:Hen3-1475_Hubble}, the high-velocity molecular outflow observed in Hen~3-1475
is well aligned with the images of [NII] emission obtained with the {\it HST}.
Interestingly, the molecular gas appears in the dark regions between the bright clumps. The clumps observed by the {\it HST} show no
significant proper motion, which has been interpreted as an indication that these structures
trace quasi-stationary shocks \citep{Fang2018}.

The projected velocity of the $^{13}$CO emission increases with distance from the star up to $\sim 240$~km~s$^{-1}$ at the emission clumps
observed the farthest away from the star (at $\sim 5\arcsec$). The ionised gas displays velocities that also increase up to the same distance, but
to a maximum of
$\sim 600$~km~s$^{-1}$ at $\sim 5\arcsec$. The ionised gas, then,
decelerate outwards by $\sim 100$~km~s$^{-1}$ up to $\sim 9\arcsec$ from the central source \citep{Bobrowsky1995}.
Hence, the velocity of the molecular gas is a factor of roughly two lower than that of the ionised gas at a similar distance from the star,
 indicating that most of the molecular material is entrained by a faster outflow.

The bright knot to the northeast
has been detected in X-rays \citep{Sahai2003b}. Strong emission of energetic radiation in the clumps of ionised gas could be the cause for
the lack of molecular emission from them. It is also possible
that molecules are destroyed by the shock, but form again after (part of) the gas cools in the post-shock region. 

The oxygen isotopic ratio we determine implies
that Hen~3-1475 is not a high-mass O-rich post-AGB star, as suggested in the literature \citep{Borkowski2001,Manteiga2011},
but had a more typical initial mass for a post-AGB star ($\lesssim 2~M_\odot$). The interpretation of this object (and similar ones) as the product
of binary interactions \citep{Borkowski2001} seems the most plausible option.

\subsection{Comparison between GLMP~950 and GLMP~953}

GLMP~950 and GLMP~953 display many similarities, but also striking differences. For instance, while the fluxes of the
$^{13}$CO, C$^{17}$O, and C$^{18}$O~\jtwoone\, lines (Table~\ref{tab:meas}) and the derived circumstellar masses (Table~\ref{tab:isot})
differ by less than a factor of two between the two sources, the continuum emission measured at $\sim 1.3$~mm (Table~\ref{tab:cont}) differs
by more than a factor of twenty. GLMP~953 shows higher values in both line and continuum emission.

Another significant difference
is that high-velocity emission ($\upsilon > 30$~km~s$^{-1}$ from the systemic velocity) is observed towards GLMP~953, but not towards GLMP~950.

The large difference in continuum emission
is puzzling, given the other similarities, the similar kinematic ages, initial masses,
circumstellar gas masses and averaged mass-loss rates derived for the two sources.
One possibility is that the emission from GLMP~953 at these wavelengths is mainly due to free-free rather than dust emission. However,
the flux density of this source at 353~GHz of 0.32~Jy \citep{Eden2017} implies a spectral index between 353 and 228~GHz of $\sim 2.3$, which does not
suggest strong free-free emission.

The number of detected lines is similar towards the two sources. However, on the one hand,
we observe lines we assign to \mbox{K$^{37}$Cl}
towards GLMP~953 (if confirmed, most likely produced in the inner and denser CSE), but not towards GLMP~950.
On the other hand, a few unidentified lines are clearly detected towards GLMP~950,
but not towards GLMP~953.

The cause of these differences is not obvious from the observations we report. While a different inclination could explain why we do not
see high-velocity emission towards GLMP~950, it does not account for the much weaker continuum emission and the slightly different molecular spectra. 
A difference in C/O ratio could potentially explain the different chemistry and stronger continuum if it leads to the formation of more dust
in GLMP~953, or of different types of dust between the two sources. { Additionally, the high-velocity emission towards GLMP~953 hints at the presence of
a companion in that system. If that is indeed the case, and the lack of high-velocity emission towards GLMP~950 indicates no companion is present, the observed
differences might be all related to GLMP~953 being in a binary system and GLMP~950 not.}

\section{Summary and conclusions}
\label{sec:conc}

We observed the \jtwoone\, lines of $^{13}$CO, C$^{17}$O, and C$^{18}$O with ALMA towards a sample of six obscured post-AGB stars:
four carbon-rich (GLMP~950, GLMP~953, AFGL~5385,
and HD~187885) and two oxygen-rich sources (Hen~3-1475 and M1-92). Emission was detected and emission-region sizes were
determined for all sources in the three lines and 1.3~mm continuum. The richness of the spectra varies strongly within the sample, with detection of
emission lines assigned by us to NaCl, SiC$_2$, SiS, K$^{37}$Cl (tentative), and HC$_3$N towards the carbon-rich sources
and SO, SO$_2$, and SiS towards the O-rich sources. The two sources
with by far the largest number of detected lines (GLMP~950 and GLMP~953)
have comparable strengths of the lines of CO isotopologues and
similar luminosities, initial masses, and kinematic ages of the C$^{18}$O~$J=2-1$ emission,
but continuum emission at 238~GHz that differs by a factor of $\sim 20$.

All but one source (GLMP~950) show high-velocity emission in the form of low-intensity wings ($\upsilon \gtrsim 30$~km~s$^{-1}$) in the $^{13}$CO~$J=2-1$ line.
The emission regions of these components show point-symmetric structure in all cases. The high-velocity
molecular outflow of Hen~3-1475 was found to be particularly remarkable, with expansion velocities $> 200$~km~s$^{-1}$ and emission
regions that are interleaved with emission from hotter ionized gas observed by the {\it HST}. The lower velocities of the molecular gas
with respect to the ionized gas at the same distance from the central source suggests that the molecules are entrained in the atomic outflow.

We inferred initial masses between 1.15 and 1.8 for the stars in our sample using the $^{17}$O/$^{18}$O isotopic ratio,
which we derived based on the observed ratios of the C$^{17}$O and C$^{18}$O lines.
We found high circumstellar masses (between $\sim 0.1$ and $\sim 0.42$~$M_\odot$)
and low kinematic ages (ranging from $\sim 600$ to $\sim 1500$~yr) corresponding to the C$^{18}$O~$J=2-1$
emission. Together,
these values imply very large recent mass-loss rates ($\gtrsim 10^{-4}~M_\odot$~yr$^{-1}$).
The high effective temperatures of AFGL~5385, HD~187885, Hen~3-1475, and M1-92 and the relatively short times since
the phases of strong mass loss imply that the central sources have evolved faster than expected for single stars.

Our results highlight how our understanding of post-AGB stars can advance using ALMA observations
of galactic sources and through a comprehensive study of their properties
(circumstellar mass, kinematic ages, and isotopic ratios).
The extent to which the population of obscured post-AGB stars are the product of single-star evolution or binary interactions remains open,
but we confirmed that their characteristics are similar to those of water fountains, which are suggested to be the product of
binary interactions. Moreover, the two O-rich sources M1-92 and Hen~3-1475 have remained rich in oxygen despite having initial masses
in the range for which becoming carbon rich is expected. 
In conclusion, observations of this type of a larger sample of objects have the potential to substantially
advance our understanding of post-AGB stars, of planetary
nebulae, and of the end of 
the AGB.

\begin{acknowledgements}
The authors thank the referee Eric Lagadec for his questions and suggestions which improved the quality of the manuscript.
The authors thank Marcelo Miller Bertolami for helpful discussions regarding the interpretation of timescales for post-AGB evolution in his models.
This paper makes use of the following ALMA data: ADS/JAO.ALMA\#2021.1.01259.S ALMA is a partnership of ESO (representing its member states), NSF (USA) and NINS (Japan), together with NRC (Canada), MOST and ASIAA (Taiwan), and KASI (Republic of Korea), in cooperation with the Republic of Chile. The Joint ALMA Observatory is operated by ESO, AUI/NRAO and NAOJ.
T.K. acknowledges support from the Swedish Research Council through grants 2019-03777. JA, VB, CSC are partially supported by I+D+i project, PID2019-105203GB-C21/C22, funded by the Spanish MICIU/AEI/10.13039/501100011033.
JFG is financially supported by grants PID2020-114461GB-I00, PID2023-146295NB-I00, and CEX2021-001131-S, funded by MCIN/AEI /10.13039/501100011033.
L.V.-P. acknowledges support from the grant PID2020-117034RJ-I00 funded by the Spanish MCIN/AEI/10.13039/501100011033.
R.S.’s contribution to the research described here was carried out at the Jet Propulsion Laboratory, California Institute of Technology,
under a contract with NASA, and funded in part by NASA via multiple ADAP and HST GO awards.
AK was supported by the Australian Research Council Centre of Excellence for All Sky Astrophysics in 3 Dimensions (ASTRO 3D),
through project number CE170100013.
MS acknowledges support from the Research Council of Norway via ESGC project (project No. 335497).
\end{acknowledgements}

\bibliographystyle{aa}
\bibliography{bibliography_2}

\begin{appendix}

\FloatBarrier
\onecolumn

\section{Additional table}

\begin{table}[h!]
\centering
\caption{Lines other than those of CO isotopologues detected towards the six post-AGB sources.}
\label{tab:lines}
\begin{tabular}{ c @{\phantom{aa}}c c c @{\phantom{aa}}c @{\phantom{aa}}c @{\phantom{aa}}c @{\phantom{aa}}c}
& & GLMP & GLMP & AFGL & HD  & Hen & M1-92 \\
Spectral lines & Freq.~[GHz] & 950 & 953 & 5385 & 187885 & 3-1475 & \\
\hline
\rowcolor{Gray}
NaCl~$v=1, J=17-16$ & 219.61498 &  & X & & & & \\ 
SO~3$\Sigma,~J=6_5-5_4$ & 219.94944 & & & & & X & X \\ 
\rowcolor{Gray}
Si$^{13}$CC~$J=17_{2,15}-17_{2,16}$ (?) & 220.598 & X & & & & & \\
 ? & 220.643 & X & & & & & \\ 
\rowcolor{Gray}
? & 220.685 & X & & & & & \\
? & 220.714 & X & X & & & & \\
\rowcolor{Gray}
? & 220.746 & X & X & & & & \\
SiC$_2$~$J=10_{0,10} - 9_{0,9}$ & 220.77369 & X & X & & & & \\
\rowcolor{Gray}
NaCl~$J=17-16$ & 221.26016 & X & X & & & & \\ 
? & 221.297 & & X & & & &  \\
\rowcolor{Gray}
? & 223.303 & X &  & & & & \\
K$^{37}$Cl~$J=30-29$ (?) & 223.78283 & & X & & & & \\ 
\rowcolor{Gray}
NaCl~$J=18-17$ & 234.25192 & X & X & & & & \\ 
SiC$_2$~$J=10_{8,2} - 9_{8,1}$ & 234.53401 & X & X & & & & \\
\rowcolor{Gray}
SiC$_2$~$J=10_{8,3} - 9_{8,2}$ & 234.53401 & X & X & & & & \\
SiS~$v=1, J=13-12$ & 234.814 &  & X & & \\
\rowcolor{Gray}
SO$_2$~$J=4_{2,2}-3_{1,3}$ & 235.15172 & & & & & X & X \\
SiC$_2$~$J=10_{6,5} - 9_{6,4}$ & 235.71301 & X & X & & \\
\rowcolor{Gray}
SiC$_2$~$J=10_{6,4} - 9_{6,3}$ & 235.71307 & X & X & & & & \\
K$^{37}$Cl~$v=2, J=32-31$~(?) & 235.766 &  & X & & & & \\
\rowcolor{Gray}
SiS~$J=13-12$~(incomplete profile) & 235.96135 & X & X & & & X & \\
SO$_2$~$J=16_{1,15}-15_{2,14}$ & 236.21669 & & & & & X & \\
\rowcolor{Gray}
HC$_3$N~$~J=26-25$ & 236.51279 & X & X & X & & &  \\
SO$_2$~$J=12_{3,9}-12_{2,10}$ & 237.06883 & & & & & X & \\
\rowcolor{Gray}
HC$_3$N~$~v_7=1, J=26-25, I=1e$ & 237.09318 & X & X & & & & \\
HC$_3$N~$~v_6=1, J=26-25, I=1f$ & 237.09318 & X & X & & & & \\
\rowcolor{Gray}
SiC$_2$~$J=10_{4,7} - 9_{4,6}$ & 237.15002 & X & X & & & & \\
SiC$_2$~$J=10_{4,7} - 9_{4,6}$ & 237.33131 & X & X & & \\
\rowcolor{Gray}
HC$_3$N~$~v_7=1, J=26-25, I=1f$ & 237.43205 & X & X & & & & \\
\end{tabular}
\end{table}

\FloatBarrier
\vspace{10cm}

\section{Velocity-integrated maps of the $^{13}$CO, C$^{17}$O, and C$^{18}$O\,\jtwoone\, lines}

\begin{figure*}[h!]
\begin{center}
 \includegraphics[width=0.9\textwidth]{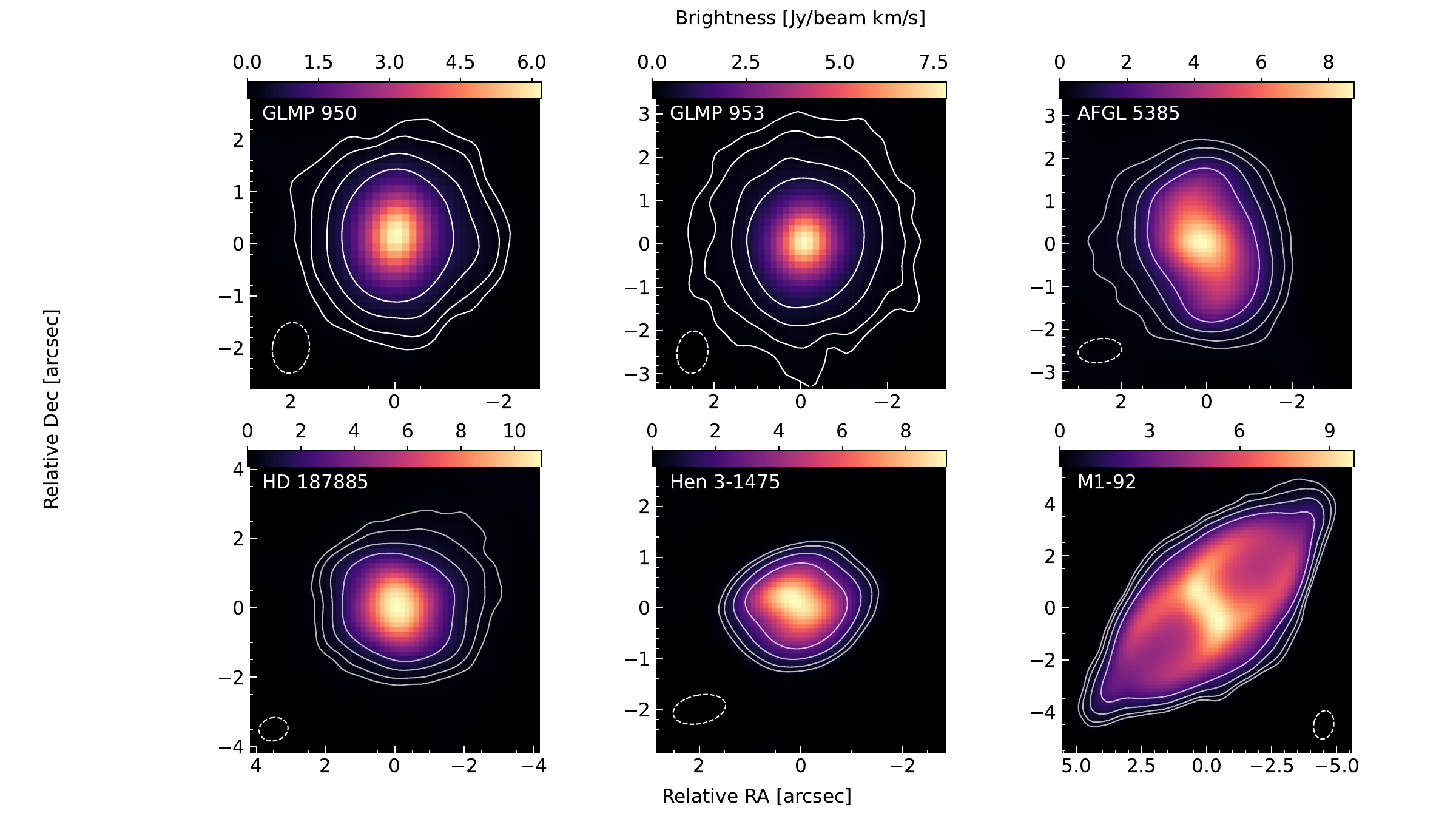}
 \caption{Brightness distribution of the $^{13}$CO\,\jtwoone\, lines integrated over the velocity interval corresponding to the full width at zero intensity of the
 C$^{18}$O\,\jtwoone\, lines. The contours show levels at 3, 5, 10, and 20 times the root mean square
 noise level of the image.}
   \label{fig:13CO}
\end{center}
\end{figure*}

\begin{figure*}[h!]
\begin{center}
 \includegraphics[width=0.9\textwidth]{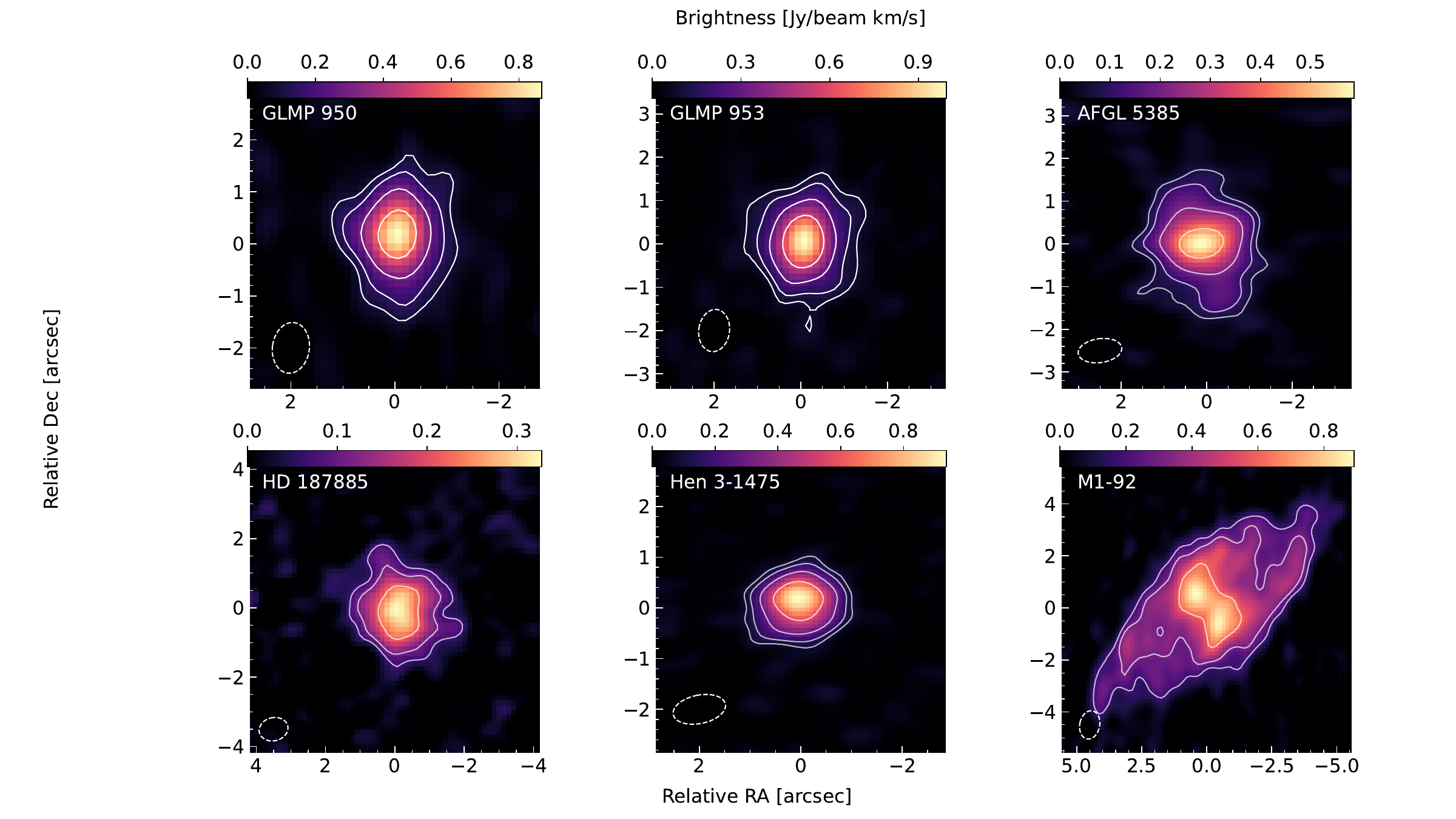}
 \caption{Brightness distribution of the C$^{17}$O\,\jtwoone\, lines integrated over their full width at zero intensity. The contours show levels at 3, 5, 10, and 20 times the root mean square
 noise level of the image.}
   \label{fig:C17O}
\end{center}
\end{figure*}

\begin{figure*}[h!]
\begin{center}
 \includegraphics[width=0.9\textwidth]{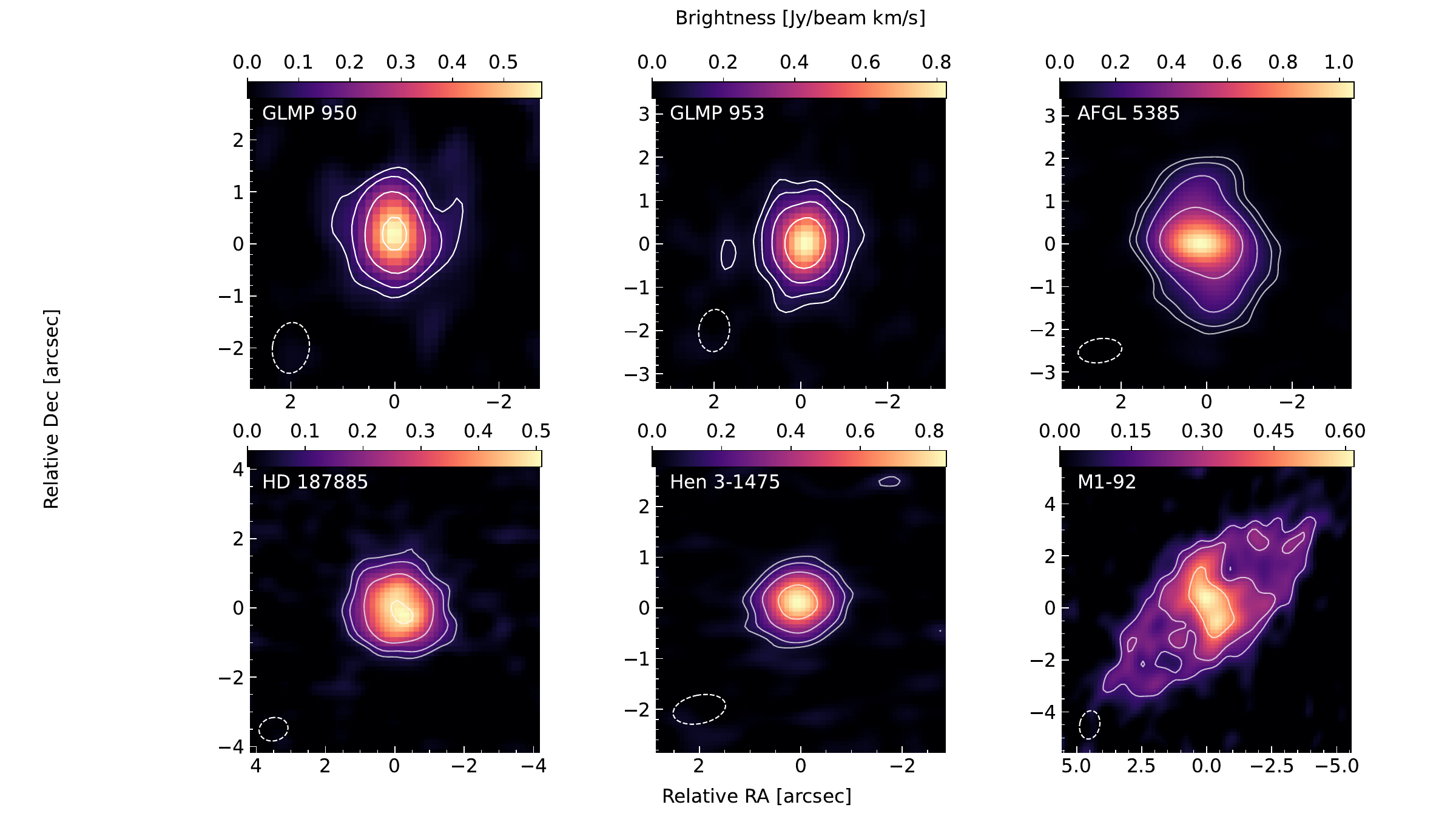}
 \caption{Brightness distribution of the C$^{18}$O\,\jtwoone\, lines integrated over their full width at zero intensity. The contours show levels at 3, 5, 10, and 20 times the root mean square
 noise level of the image.}
   \label{fig:C18O}
\end{center}
\end{figure*}

\FloatBarrier

\section{Spitzer images of GLMP~950 and GLMP~953}

\begin{figure*}[h!]
\begin{center}
 \includegraphics[width=0.48\textwidth]{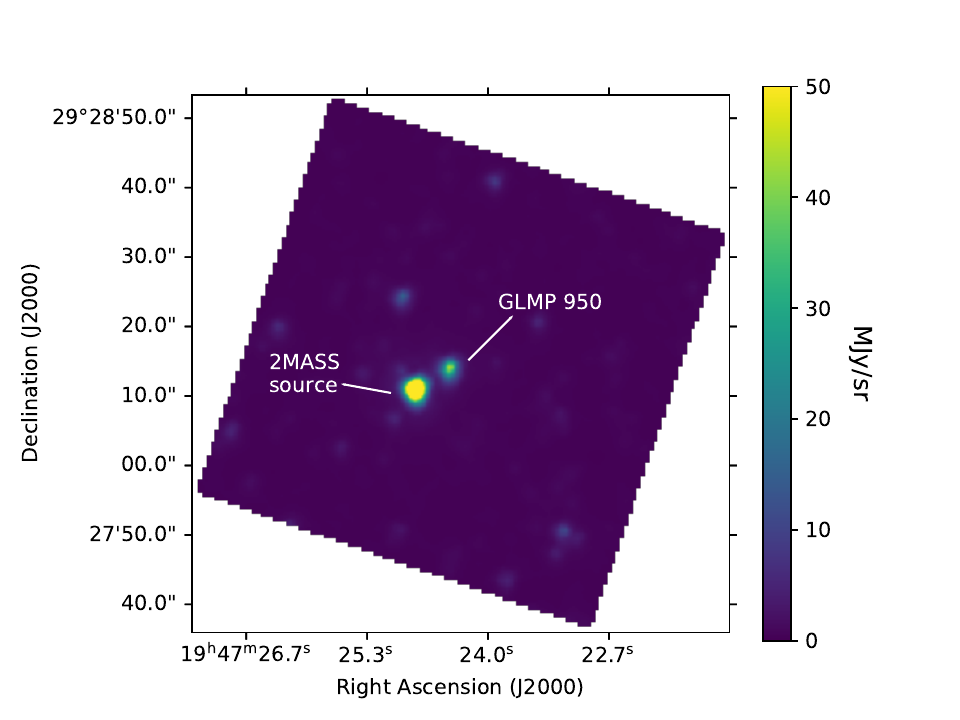}
 \includegraphics[width=0.48\textwidth]{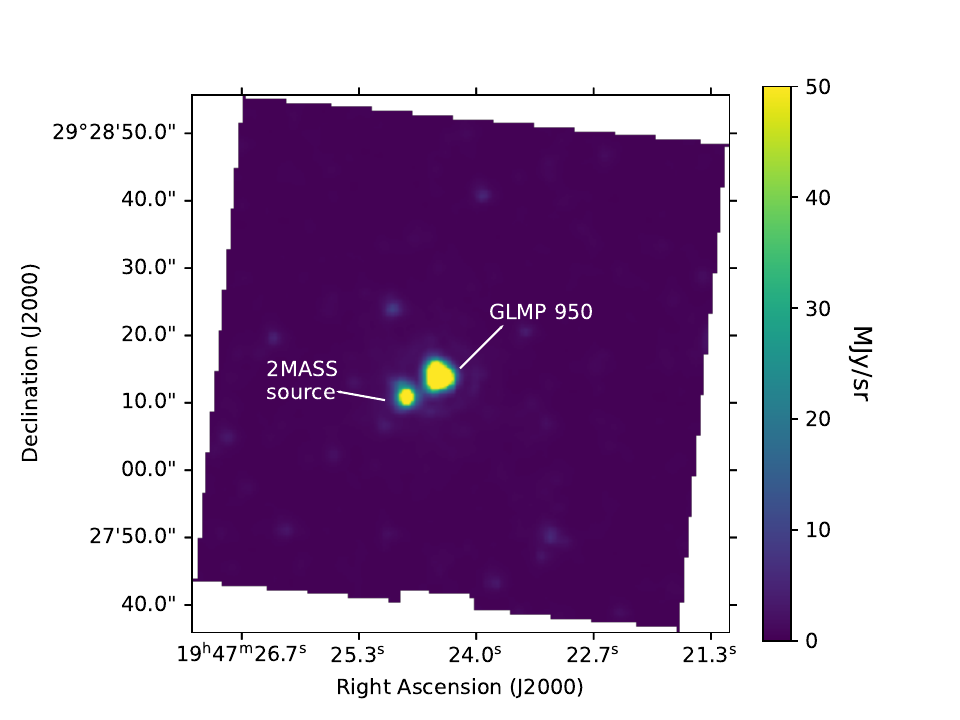} 
 \caption{Cut-out of Spitzer/IRAC images at 3.6~$\mu$m (left) and at 4.5~$\mu$m (right) towards GLMP~950 and 2MASS~J19472480+2928108.}
   \label{fig:Spitzer_GLMP950}
\end{center}
\end{figure*}

\begin{figure*}[h!]
\begin{center}
 \includegraphics[width=0.49\textwidth]{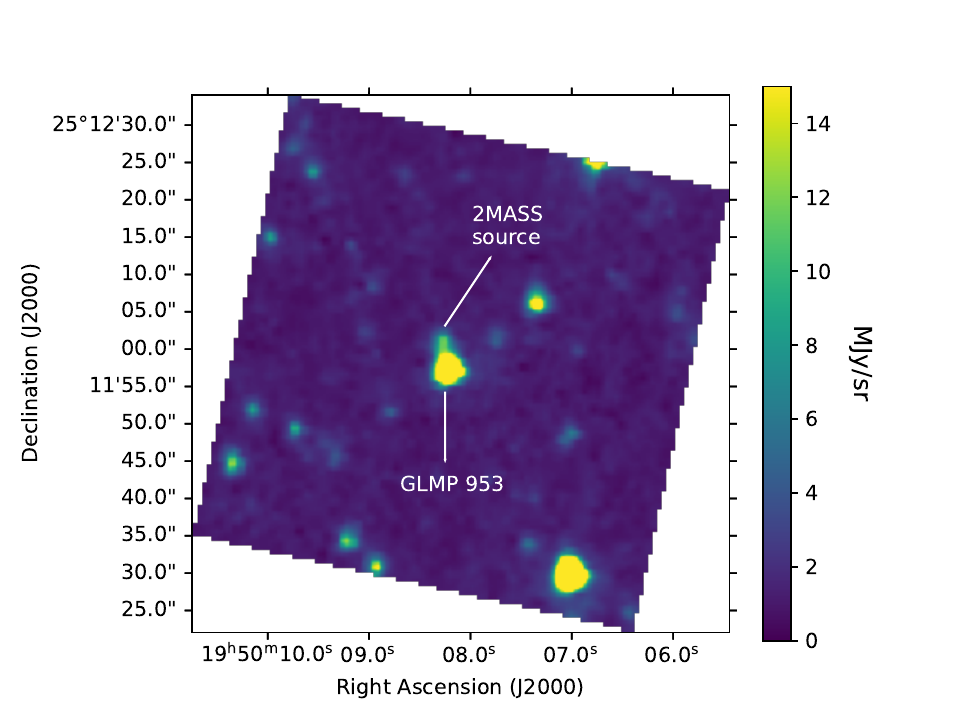} 
 \includegraphics[width=0.49\textwidth]{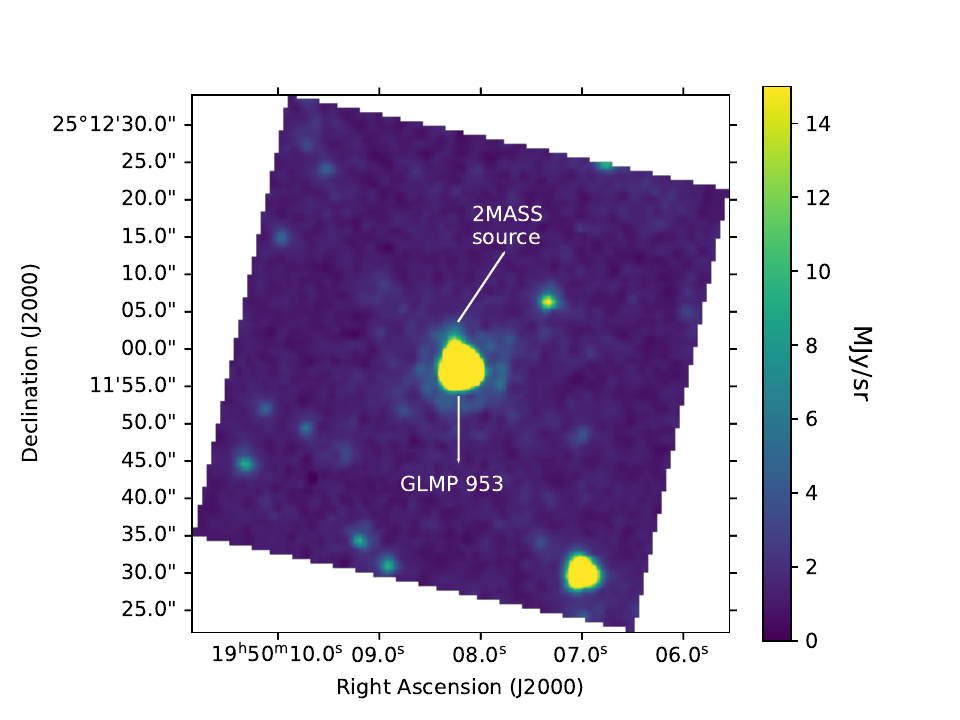} 
 \caption{Cut-out of Spitzer/IRAC images at 3.6~$\mu$m (left) and 4.5~$\mu$m (right) towards GLMP~953 and 2MASS~J19500826+2512008.}
   \label{fig:Spitzer_GLMP953}
\end{center}
\end{figure*}

\FloatBarrier

\section{Continuum image of AFGL~5385 with additional source}

\begin{figure*}[h!]
\begin{center}
 \includegraphics[width=0.49\textwidth]{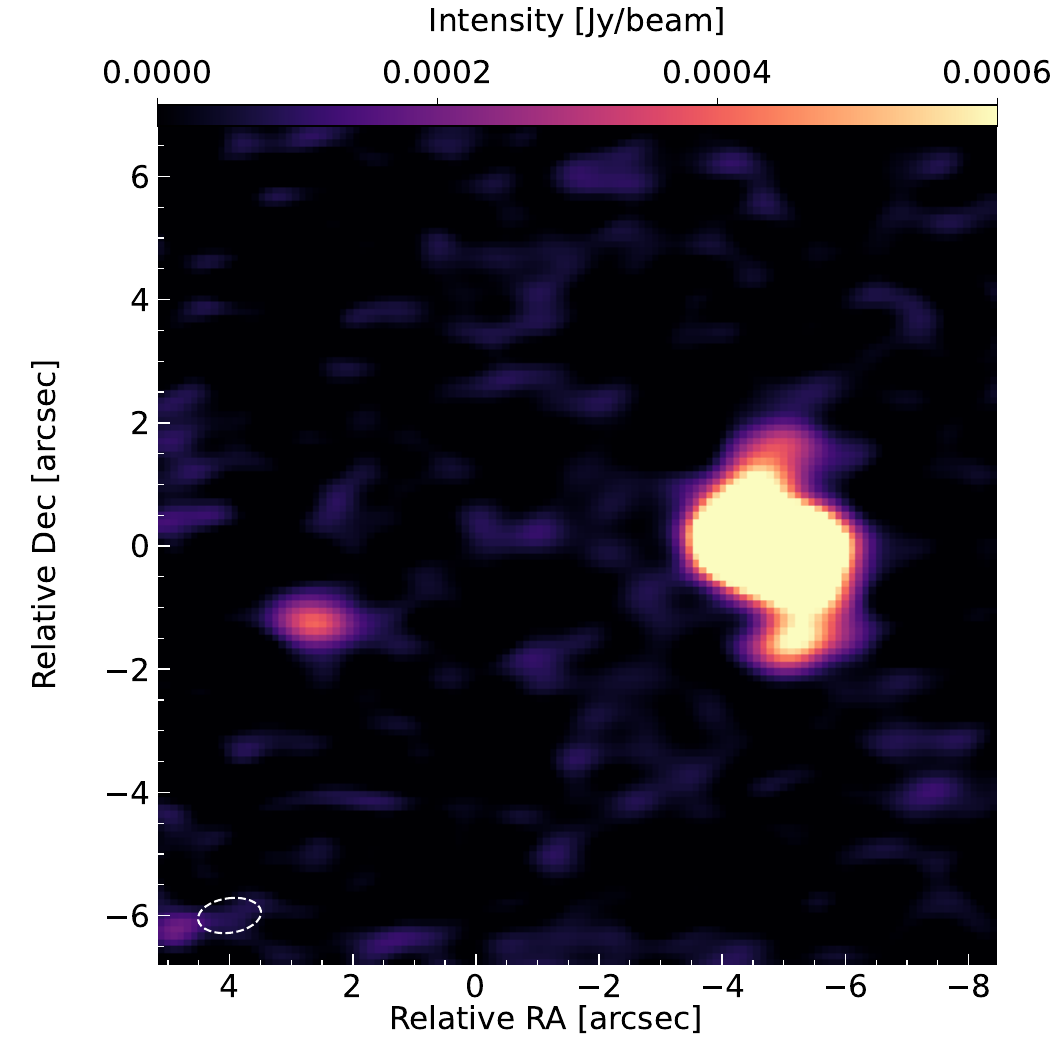} 
 \caption{Continuum image of AFGL~5385 including a larger region than Fig.~\ref{fig:AFGL_highVel} to show the second detected source to the east.}
   \label{fig:AFGL_cont}
\end{center}
\end{figure*}

\end{appendix}
\end{document}